\documentclass[aps,pra,reprint,superscriptaddress,nofootinbib]{revtex4-2}

\usepackage{amsmath,amssymb,bm,graphicx,hyperref}
\hypersetup{colorlinks=true,linkcolor=blue,citecolor=blue,urlcolor=blue}
\usepackage{epstopdf}
\newcommand{\ii}{\mathrm{i}}
\newcommand{\ee}{\mathrm{e}}

\newcommand{\hc}{\mathrm{H.c.}}
\newcommand{\ket}[1]{|#1\rangle}

\newcommand{\I}{\mathbb{I}}
\newcommand{\maybegraphics}[2][0.95\columnwidth]{\IfFileExists{#2}{\includegraphics[width=#1]{#2}}{\rule{0pt}{0.28\textheight}\rule{#1}{0pt}}}

\begin{document}
	
	\title{Non-Hermitian scattering in SSH superconducting waveguides: exact Green-function reduction and dimerization-sensitive
		microwave functionalities}
	\author{J. Zhou}
	\affiliation{College of Physics and Materials Science, Tianjin Normal University, Tianjin
		300387, China}
		
	\author{X. X. Zhang}
	\affiliation{College of Physics and Materials Science, Tianjin Normal University, Tianjin
		300387, China}
		
	\author{X. Z. Zhang}
	\email{zhangxz@tjnu.edu.cn}
	\affiliation{College of Physics and Materials Science, Tianjin Normal University, Tianjin
		300387, China}
	\affiliation{Interdisciplinary center, Tianjin Normal University, Tianjin
			300387, China}

	\begin{abstract}
		We formulate an exact Green-function theory for non-Hermitian single-microwave-photon scattering by finite superconducting circuit subsystems embedded in an SSH waveguide. The structured SSH environment is integrated out exactly and enters the local scattering problem as an energy-dependent matrix self-energy, reducing the full open system to a finite-dimensional effective non-Hermitian Hamiltonian. This reduction places scattering amplitudes, exceptional-point diagnostics, coherent-perfect-absorption conditions, and lasing thresholds within one unified framework. Within this approach we analyze two superconducting devices. A flux-controlled two-qubit interferometric scatterer exhibits a broad bright branch and a narrow quasi-dark branch whose interference is reshaped by the SSH environment and changes qualitatively across the two dimerizations. A mediator-assisted two-qubit scatterer generates an additional energy-dependent complex coupling, reorganizes the dressed spectrum, and produces clearer dimerization-sensitive transparency-versus-absorption windows together with a pronounced separation between zero-like and pole-like scattering branches. In the active regime, near-exceptional-point hybridization enhances the pole-dominated response while deepening the singular-value valley associated with near-coherent perfect absorption. These results show how structured topological waveguides can be used not only to host scattering, but also to design non-Hermitian superconducting microwave functionalities.
	\end{abstract}
	
	\maketitle
	
	\section{Introduction}
	
	Waveguide quantum electrodynamics provides a direct setting for resolving how localized quantum emitters hybridize, interfere, and exchange dissipation with itinerant photons at the single-photon level. In superconducting platforms this problem is particularly flexible, because artificial atoms, microwave resonators, tunable couplers, and engineered loss or gain channels can all be controlled in situ.~\cite{houckOnchip2012,zhengWaveguideQEDBasedPhotonicQuantum2012,blais2020quantum,mahmoodianDynamicsManyBodyPhoton2020,kannanGeneratingSpatiallyEntangled2020,kannanWaveguideQuantumElectrodynamics2020,kimQuantumElectrodynamicsTopological2021,shiFastAtomphotonEntangling2022,vegaQubitphotonBoundStates2021,jinTopologicalPrethermalStrong2025,hauffChiralQuantumOptics2022,davoodiBoundStatesQuantum2025} As a result, scattering is not merely a convenient probe of the device response, but a direct way to access open-system mode structure, interference, and environment-induced dressing~\cite{shenCoherentSinglePhoton2005,zhouControllableScatteringSingle2008,roy2017colloquium,yinSinglephotonScatteringGiantmolecule2022,zhuSinglephotonScatteringGiantatom2025,luTopologicalPhotonics2014,zhaoSinglephotonScatteringBound2020,zouTunableSinglePhotonScattering2022,liScatteringStatesOnedimensional2025,nieTopologyEnhancedNonreciprocalScattering2021,RSinglephotonscattering2025}.
	
	When the waveguide itself is structured, the scattering problem acquires an additional layer of organization. Among one-dimensional topological waveguides, the Su--Schrieffer--Heeger (SSH) lattice provides a natural and experimentally meaningful setting~\cite{suSolitonsPolyacetylene1979,caceres-aravenaExperimentalObservationEdge2022}. Its dimerized bipartite structure supplies two sublattice-resolved contact channels, a controllable topological transition, and a strongly energy-dependent local response near the band edges and in the gap~\cite{wenColloquiumZooQuantumtopological2017,rechtsmanPhotonicFloquetTopological2013}. Once a finite superconducting circuit subsystem is attached locally to such a waveguide, the waveguide no longer acts as a featureless reservoir. Instead, it feeds back onto the local circuit through an energy-dependent matrix structure that depends on both the SSH band geometry and the sublattice composition of the contact region. The resulting scattering problem is therefore qualitatively different from the corresponding one in a uniform transmission line.
	
	Recent years have witnessed growing interest in single-photon transport in structured waveguides, giant-atom waveguide quantum electrodynamics (QED) systems, and topological photonic environments. Significant progress has been made in understanding nonlocal coupling effects, bound-state formation, interference-induced transparency, and non-Hermitian transport phenomena through a variety of theoretical frameworks and scattering geometries ~\cite{PhysRevA.104.033710, PhysRevA.104.063712, chen2022nonreciprocal, PhysRevLett.128.223602, cheng2023single,PhysRevA.108.043709, PhysRevA.109.063703, PhysRevA.109.053720, PhysRevA.109.063708, PhysRevA.110.053716, PhysRevA.111.023712, 99nn-9wxs}.		
	Despite these advances, most existing studies focus on specific emitter configurations, particular coupling architectures, or model-dependent scattering scenarios. A general framework that systematically separates the influence of a structured topological environment from the internal design of a finite non-Hermitian superconducting circuit remains highly desirable.
	This observation motivates the Green-function approach developed in the present work. The main point is not simply to compute reflection and transmission amplitudes in another SSH-based geometry, but to reformulate the full open scattering problem in a way that makes the role of the structured environment explicit. By integrating out the extended SSH waveguide exactly, the waveguide is converted into an energy-resolved matrix self-energy acting on a finite-dimensional non-Hermitian superconducting circuit subsystem~\cite{benderComplexExtensionQuantum2002,benderMakingSenseNonHermitian2007,okumaTopologicalOriginNonHermitian2020,linTopologicalNonHermitianSkin2023,liObservationDynamicNonHermitian2024,liEffectiveHamiltonianPhotonic2022,gongTopologicalPhasesNonHermitian2018,lieuTopologicalPhasesNonHermitian2018,esakiEdgeStatesTopological2011,wuFloquetTopologicalPhases2020,el-ganainyNonHermitianPhysicsPT2018,banerjeeNonHermitianTopologicalPhases2023}. The entire scattering problem is then reduced to a finite-dimensional effective non-Hermitian Hamiltonian in which poles, zeros, mode hybridization, exceptional-point diagnostics, coherent-perfect-absorption conditions, and lasing thresholds can be analyzed within one common language~\cite{wangCoherentPerfectAbsorption2021,chenExceptionalPointsEnhance2017,miriExceptionalPointsOptics2019}.
	
	To our knowledge, an explicit Green-function reduction of this form for non-Hermitian superconducting scattering centers embedded in an SSH waveguide has not been formulated in this way. This formulation is useful for two related reasons. On the formal side, it separates the role of the extended topological environment from the design of the local circuit and yields compact expressions for the effective Hamiltonian, the $T$ matrix, and the two-port scattering matrix. On the physical side, it makes clear which part of the observed response is imposed by the SSH waveguide and which part is controlled by the internal structure of the local superconducting device. This separation is particularly valuable once gain, loss, and internal mode conversion are introduced, because the same framework can be used to discuss passive and active operating regimes on equal footing.

\begin{figure*}[t]
		\centering
		\maybegraphics[0.9\textwidth]{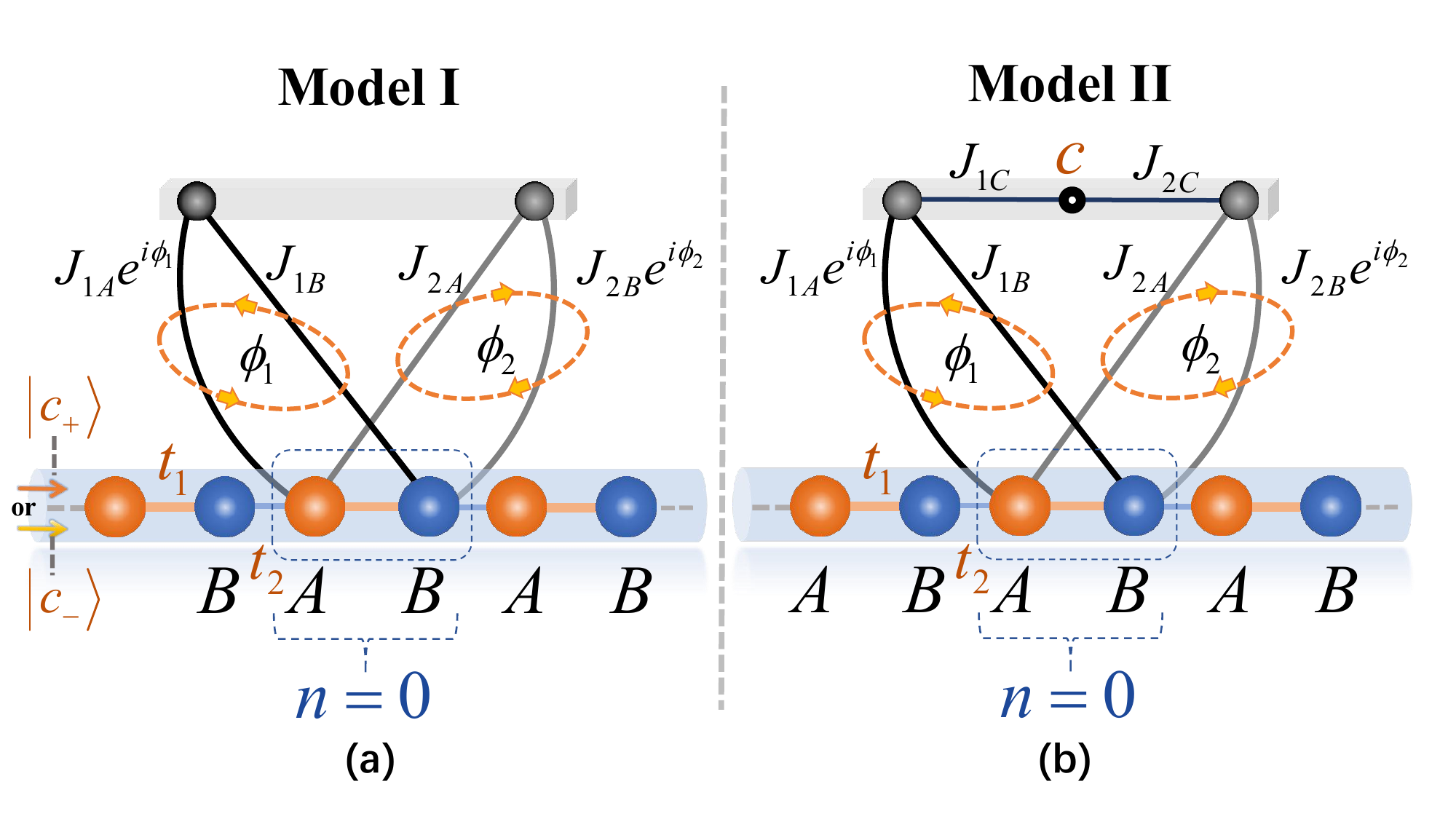}
		\caption{Unified superconducting-circuit schematics of the two non-Hermitian SSH scattering devices studied in this work. Panel (a) shows Model I, a parallel two-qubit interferometric scatterer in which both qubits couple directly to the $A$ and $B$ resonators of the same SSH unit cell and enclose a synthetic Aharonov--Bohm flux. In the symmetric geometry the natural contact combinations $|c_{\pm}\rangle=(|A\rangle\pm|B\rangle)/\sqrt2$ and the qubit molecular modes $|q_{\pm}\rangle=(|e_1\rangle\pm|e_2\rangle)/\sqrt2$ jointly organize the scattering into bright and dark molecular channels. Panel (b) shows Model II, where the two qubits are additionally coupled through an auxiliary internal mode $c$. The mediator does not couple directly to the SSH waveguide but generates an energy-dependent effective interaction between the qubits, thereby creating a mediator-dark branch and a mediator-bright polaritonic sector before the SSH contact channels further dress the scattering response.}
		\label{fig:models_schematic}
	\end{figure*}
	
	To illustrate the power of the approach,
	we apply this formulation to two superconducting devices that are locally embedded in the central unit cell of an SSH waveguide. In Model~I, two superconducting qubits form a flux-controlled interferometric scatterer coupled in parallel to the $A$ and $B$ sites of the same SSH cell. In this geometry the structured SSH environment separates a broad bright scattering branch from a narrow quasi-dark branch, and the resulting interference is reshaped across the two dimerizations. In Model~II, the two qubits are further coupled to an auxiliary internal mode. After exact elimination of that mode, the local problem acquires an additional energy-dependent complex interqubit coupling, which reorganizes the dressed spectrum and produces a clearer hierarchy of bright and quasi-dark scattering channels. This second geometry makes it possible to expose dimerization-sensitive transparency-versus-absorption windows more clearly, and also to separate more sharply the branches that dominate zero-like and pole-like scattering responses. The physical questions addressed in the two models are therefore related but not identical. Model~I isolates how the SSH environment itself reorganizes local non-Hermitian scattering through sublattice-resolved dressing and flux-controlled interference. Model~II shows how this SSH-induced structure is further reshaped by an internal mediator, leading to a more pronounced distinction between broad and narrow dressed branches, clearer Fano- and EIT-like line-shape conversion, and a more transparent route toward the analysis of near-exceptional-point, near-CPA, and pole-dominated scattering regimes.
	
	The rest of the paper is organized as follows. Section~\ref{sec:general} develops the general Green-function formulation for an SSH superconducting waveguide locally coupled to a finite non-Hermitian circuit subsystem. Section~\ref{sec:modelI} applies the theory to the interferometric two-qubit scatterer and derives compact expressions for the self-energy, effective Hamiltonian, and scattering amplitudes of Model~I. Section~\ref{sec:modelII} develops the mediator-assisted geometry, from the exact Schur-complement reduction to the passive and active scattering responses of Model~II. Section~\ref{sec:devices} summarizes the resulting implications for non-Hermitian superconducting microwave functionalities.
	
	\section{General Green-function formulation}
	\label{sec:general}
	
	\subsection{SSH superconducting circuit waveguide}
	
	We model the dimerized superconducting microwave lattice by the SSH Hamiltonian~\cite{caceres-aravenaExperimentalObservationEdge2022}
	\begin{equation}
		H_{\mathrm{wg}}=\sum_n\Big(t_1 a_n^{\dagger}b_n+t_2 a_{n+1}^{\dagger}b_n+\hc\Big),
		\label{eq:Hwg}
	\end{equation}
	where $a_n^{\dagger}$ and $b_n^{\dagger}$ create a single microwave excitation on the $A$ and $B$ resonators of unit cell $n$. The couplings $t_1$ and $t_2$ are controlled by superconducting circuit couplers and may be parametrized as $t_1=t(1+\delta)$ and $t_2=t(1-\delta)$, where $\delta$ is the dimerization parameter.
	
	After Fourier transformation,
	\begin{equation}
		a_n=\frac{1}{\sqrt{N}}\sum_k a_k\ee^{\ii kn},\qquad b_n=\frac{1}{\sqrt{N}}\sum_k b_k\ee^{\ii kn},
	\end{equation}
	we obtain
	\begin{equation}
		H_{\mathrm{wg}}=\sum_k
		\begin{pmatrix}
			a_k^{\dagger} & b_k^{\dagger}
		\end{pmatrix}
		H_0(k)
		\begin{pmatrix}
			a_k\\ b_k
		\end{pmatrix},
	\end{equation}
	with
	\begin{equation}
		H_0(k)=
		\begin{pmatrix}
			0 & h(k)\\
			h^*(k) & 0
		\end{pmatrix},
		\qquad h(k)=t_1+t_2\ee^{-\ii k}.
		\label{eq:H0k_main}
	\end{equation}
	Writing $h(k)=d_x(k)-\ii d_y(k)=|h(k)|\ee^{-\ii\theta_k}$ gives
	\begin{equation}
		d_x(k)=t_1+t_2\cos k,
		\qquad d_y(k)=t_2\sin k,
	\end{equation}
	and the band energies
	\begin{equation}
		E_s(k)=s\,\varepsilon_k,~~~ \varepsilon_k=\sqrt{t_1^2+t_2^2+2t_1t_2\cos k},
		~~~s=\pm.
		\label{eq:dispersion_main}
	\end{equation}
	A convenient Bloch spinor basis is
	\begin{equation}
		u_s(k)=\frac{1}{\sqrt{2}}
		\begin{pmatrix}
			1\\ s\ee^{\ii\theta_k}
		\end{pmatrix},
		\qquad H_0(k)\,u_s(k)=E_s(k)\,u_s(k).
		\label{eq:spinor_main}
	\end{equation}
	The group velocity in band $s$ is
	\begin{equation}
		v_s(k)=\left|\frac{\partial E_s(k)}{\partial k}\right|=\frac{|t_1t_2\sin k|}{\varepsilon_k}.
		\label{eq:velocity_main}
	\end{equation}
	
	\subsection{Local and nonlocal SSH Green functions}
	
	The retarded Green function of the clean SSH waveguide is
	\begin{equation}
		g^r(E)=\frac{1}{E+\ii0^+-H_{\mathrm{wg}}}.
	\end{equation}
	Because the scatterer couples only to the central unit cell $n=0$, the key quantity is the local two-by-two Green function in the contact subspace,
	\begin{equation}
		g_{00}^r(E)=\int_{-\pi}^{\pi}\frac{dk}{2\pi}\,\frac{1}{E+\ii0^+-H_0(k)}.
		\label{eq:g00_main}
	\end{equation}
	Using $H_0(k)^2=\varepsilon_k^2\I$, one finds
	\begin{equation}
		g_{00}^r(E)=g_0(E)\I+g_x(E)\sigma_x,
		\label{eq:g00decomp_main}
	\end{equation}
	where
	\begin{equation}
		g_0(E)=\int_{-\pi}^{\pi}\frac{dk}{2\pi}\,\frac{E}{(E+\ii0^+)^2-\varepsilon_k^2},
		\label{eq:g0_main}
	\end{equation}
	\begin{equation}
		g_x(E)=\int_{-\pi}^{\pi}\frac{dk}{2\pi}\,\frac{t_1+t_2\cos k}{(E+\ii0^+)^2-\varepsilon_k^2}.
		\label{eq:gx_main}
	\end{equation}
	The absence of a $\sigma_y$ term reflects the cancellation of opposite momenta in a local return amplitude.
	
	The nonlocal propagator from the contact cell to cell $n$ is
	\begin{equation}
		g_{n0}^r(E)=\int_{-\pi}^{\pi}\frac{dk}{2\pi}\,\ee^{\ii kn}\,\frac{1}{E+\ii0^+-H_0(k)}.
		\label{eq:gn0_main}
	\end{equation}
	For real $E$ inside a propagating band, the large-distance asymptotics is controlled by the on-shell poles and takes the form
	\begin{equation}
		g_{n0}^r(E)\sim -\frac{\ii}{v_s(k)}\,\ee^{\ii kn}\,u_s(k)u_s^{\dagger}(k),
		\qquad n\to +\infty,
		\label{eq:asympright_main}
	\end{equation}
	\begin{equation}
		g_{n0}^r(E)\sim -\frac{\ii}{v_s(k)}\,\ee^{-\ii k|n|}\,u_s(-k)u_s^{\dagger}(-k),
		~~~ n\to -\infty.
		\label{eq:asympleft_main}
	\end{equation}
	These formulas are the Green-function counterparts of right-going and left-going SSH Bloch waves.
	
	\subsection{Finite-dimensional non-Hermitian scatterer}
	
	We now attach a finite superconducting circuit subsystem to the contact cell. In the single-excitation sector the internal amplitudes are collected in an $M$-component vector $X$, and the scatterer Hamiltonian is an $M\times M$ matrix $H_{\mathrm{sc}}$. The coupling between the SSH contact spinor $\Psi_0=(a_0,b_0)^T$ and the scatterer is encoded in a $2\times M$ matrix $W$ through
	\begin{equation}
		H_{\mathrm{cpl}}=\Psi_0^{\dagger}WX+X^{\dagger}W^{\dagger}\Psi_0.
	\end{equation}
	The internal frequencies may be complex, $\tilde\omega_\mu=\omega_\mu+\ii\gamma_\mu$, where $\gamma_\mu<0$ represents net loss and $\gamma_\mu>0$ represents effective gain.
	
	Because the scatterer is finite dimensional, its internal amplitudes can be eliminated exactly. This yields an energy-dependent effective potential acting on the SSH contact subspace,
	\begin{equation}
		V_{\mathrm{eff}}(E)=W(E-H_{\mathrm{sc}})^{-1}W^{\dagger}.
		\label{eq:Veff_main}
	\end{equation}
	The dressed local Green function is
	\begin{equation}
		G_{00}^r(E)=\Big[(g_{00}^r)^{-1}-V_{\mathrm{eff}}(E)\Big]^{-1},
		\label{eq:G00_main}
	\end{equation}
	and the corresponding $T$ matrix is
	\begin{equation}
		T(E)=W\Big[E-H_{\mathrm{sc}}-\Sigma^r(E)\Big]^{-1}W^{\dagger},
		\label{eq:T_main}
	\end{equation}
	with the SSH-induced self-energy
	\begin{equation}
		\Sigma^r(E)=W^{\dagger}g_{00}^r(E)W.
		\label{eq:Sigma_main}
	\end{equation}
	It is therefore natural to define the energy-dependent effective non-Hermitian Hamiltonian
	\begin{equation}
		H_{\mathrm{eff}}(E)=H_{\mathrm{sc}}+\Sigma^r(E).
		\label{eq:Heff_main}
	\end{equation}
	All scattering observables are controlled by this finite-dimensional matrix.
	
	\subsection{Scattering states and the two-port scattering matrix}
	
	For a left-incident Bloch wave in band $s$,
	\begin{equation}
		\ket{\Phi_L^{\mathrm{in}}}=\sum_n\ee^{\ii kn}u_s(k)\ket{n},
		\qquad E=E_s(k),
	\end{equation}
	the exact outgoing state obeys the Lippmann--Schwinger equation
	\begin{equation}
		\ket{\Psi_L^{(+)}}=\ket{\Phi_L^{\mathrm{in}}}+g^r(E)T(E)\ket{\Phi_L^{\mathrm{in}}}.
		\label{eq:LS_main}
	\end{equation}
	Using Eqs.~(\ref{eq:asympright_main}) and (\ref{eq:asympleft_main}), we obtain the left-incidence transmission and reflection amplitudes
	\begin{equation}
		t_L^{(s)}(E)=1-\frac{\ii}{v_s(k)}u_s^{\dagger}(k)T(E)u_s(k),
		\label{eq:tL_main}
	\end{equation}
	\begin{equation}
		r_L^{(s)}(E)=-\frac{\ii}{v_s(k)}u_s^{\dagger}(-k)T(E)u_s(k).
		\label{eq:rL_main}
	\end{equation}
	For right incidence,
	\begin{equation}
		\ket{\Phi_R^{\mathrm{in}}}=\sum_n\ee^{-\ii kn}u_s(-k)\ket{n},
	\end{equation}
	one similarly finds
	\begin{equation}
		t_R^{(s)}(E)=1-\frac{\ii}{v_s(k)}u_s^{\dagger}(-k)T(E)u_s(-k),
		\label{eq:tR_main}
	\end{equation}
	\begin{equation}
		r_R^{(s)}(E)=-\frac{\ii}{v_s(k)}u_s^{\dagger}(k)T(E)u_s(-k).
		\label{eq:rR_main}
	\end{equation}
	The two-port scattering matrix is therefore
	\begin{equation}
		S(E)=
		\begin{pmatrix}
			r_L(E) & t_R(E)\\
			t_L(E) & r_R(E)
		\end{pmatrix}.
		\label{eq:S_main}
	\end{equation}
	The measurable transmittance and reflectance are
	\begin{equation}
		\mathcal{T}_{L,R}=|t_{L,R}|^2,
		\qquad
		\mathcal{R}_{L,R}=|r_{L,R}|^2,
		\label{eq:RT_main}
	\end{equation}
	and the deviation from flux conservation is quantified by
	\begin{equation}
		\mathcal{A}_{L,R}=1-\mathcal{R}_{L,R}-\mathcal{T}_{L,R},
		\label{eq:A_main}
	\end{equation}
	with $\mathcal{A}>0$ indicating net absorption and $\mathcal{A}<0$ indicating net amplification.
	
	The pole condition of the scattering matrix is inherited from the finite-dimensional resolvent,
	\begin{equation}
		\det\Big[E-H_{\mathrm{eff}}(E)\Big]=0.
		\label{eq:pole_main}
	\end{equation}
	For a two-port device, coherent perfect absorption corresponds to a zero eigenvalue of $S(E)$ on the real axis, or equivalently
	\begin{equation}
		\det S(E)=0,
		\qquad E\in\mathbb{R}.
		\label{eq:CPA_main}
	\end{equation}
	A lasing threshold occurs when a pole of $S(E)$ reaches the real axis. Exceptional points are branch-point degeneracies of $H_{\mathrm{eff}}(E)$ and are detected by the coalescence of both eigenvalues and eigenvectors.

	\section{Model I: parallel two-qubit interferometric scatterer}
	\label{sec:modelI}
	
	\subsection{Microscopic Hamiltonian and coupling matrix}
	
	The first concrete device is the parallel two-qubit interferometer shown in Fig.~\ref{fig:models_schematic}(a). Two superconducting qubits are both attached to the $A$ and $B$ resonators of the same SSH unit cell and enclose a synthetic Aharonov--Bohm flux. In the single-excitation sector the internal basis is $X=(\ket{e_1},\ket{e_2})^T$, and the bare non-Hermitian qubit Hamiltonian is
	\begin{equation}
		H_{\mathrm{sc}}^{(\mathrm{I})}=
		\begin{pmatrix}
			\tilde\omega_1 & 0\\
			0 & \tilde\omega_2
		\end{pmatrix},
		\qquad
		\tilde\omega_j=\omega_j+\ii\gamma_j .
		\label{eq:HscI}
	\end{equation}
	Here $\gamma_j<0$ represents net loss of qubit $j$, whereas $\gamma_j>0$ models effective gain generated by reservoir engineering or parametric pumping.
	
	The four qubit--waveguide couplings are
	\begin{align}
		H_{\mathrm{cpl}}^{(\mathrm{I})}&=
		\Big(J_{1A}\ee^{\ii\phi_1}a_0^{\dagger}+J_{1B}b_0^{\dagger}\Big)\sigma_1^-\nonumber\\
		&+\Big(J_{2A}a_0^{\dagger}+J_{2B}\ee^{\ii\phi_2}b_0^{\dagger}\Big)\sigma_2^-+\hc ,
		\label{eq:HcplI}
	\end{align}
	so that the contact matrix is
	\begin{equation}
		W_{\mathrm{I}}=
		\begin{pmatrix}
			J_{1A}\ee^{\ii\phi_1} & J_{2A}\\
			J_{1B} & J_{2B}\ee^{\ii\phi_2}
		\end{pmatrix}.
		\label{eq:WI}
	\end{equation}
	The total synthetic flux piercing the interferometric loop is $\Phi=\phi_1+\phi_2$ modulo a gauge convention. In a superconducting realization $\phi_{1,2}$ are controlled by flux-biased couplers or parametrically modulated coupling elements, so the interferometer can be tuned \emph{in situ} from a fully constructive to a fully destructive geometry.

	Figure~\ref{fig:models_schematic} is the device-level roadmap for the rest of the paper. Panel~(a) isolates the logic of Model~I: both qubits talk directly to the same SSH contact cell, so the central control knob is interferometric. The synthetic flux first decides which molecular qubit combination is bright and which one is dark or quasi-dark, and the SSH environment then dresses those two contact-selected branches. Panel~(b) adds one more internal stage. Before the waveguide enters, the auxiliary mode $c$ already separates the internal spectrum into a mediator-bright polaritonic sector and a mediator-dark qubit branch; only after that internal selection do the SSH contact channels decide how strongly each branch appears in transport.
	
	Placed side by side, the two schematics also explain why the two models are not just minor variants of one another. Model~I is a one-stage bright/dark interferometer dressed by the SSH waveguide. Model~II is a two-stage device: the mediator first reorganizes the local spectrum, and the SSH ports then read out that reorganized spectrum. This extra hierarchy is the reason Model~II supports a broader range of dressed line shapes, including sharper branch selectivity, mediator-induced Fano structures, and a cleaner separation between zero-dominated and pole-dominated responses.
	
	\subsection{Exact SSH self-energy and effective Hamiltonian}
	
	Using $g_{00}^r(E)=g_0(E)\I+g_x(E)\sigma_x$, the exact SSH-induced self-energy is
	\begin{equation}
		\Sigma_{\mathrm{I}}^r(E)=W_{\mathrm{I}}^{\dagger}g_{00}^r(E)W_{\mathrm{I}}.
		\label{eq:SigmaI_def}
	\end{equation}
	Its matrix elements are
	\begin{align}
		\Sigma_{11}^{(\mathrm{I})}(E)&=g_0\!\left(J_{1A}^2+J_{1B}^2\right)+2g_x J_{1A}J_{1B}\cos\phi_1 ,
		\label{eq:sigma11I}\\
		\Sigma_{22}^{(\mathrm{I})}(E)&=g_0\!\left(J_{2A}^2+J_{2B}^2\right)+2g_x J_{2A}J_{2B}\cos\phi_2 ,
		\label{eq:sigma22I}\\
		\Sigma_{12}^{(\mathrm{I})}(E)&=g_0\!\left(J_{1A}J_{2A}\ee^{-\ii\phi_1}+J_{1B}J_{2B}\ee^{\ii\phi_2}\right)\nonumber\\
		&\quad +g_x\!\left(J_{1A}J_{2B}\ee^{\ii(\phi_2-\phi_1)}+J_{1B}J_{2A}\right),
		\label{eq:sigma12I}\\
		\Sigma_{21}^{(\mathrm{I})}(E)&=g_0\!\left(J_{1A}J_{2A}\ee^{\ii\phi_1}+J_{1B}J_{2B}\ee^{-\ii\phi_2}\right)\nonumber\\
		&\quad +g_x\!\left(J_{1A}J_{2B}\ee^{-\ii(\phi_2-\phi_1)}+J_{1B}J_{2A}\right).
		\label{eq:sigma21I}
	\end{align}
	The effective Hamiltonian entering all scattering observables is therefore
	\begin{equation}
		H_{\mathrm{eff}}^{(\mathrm{I})}(E)=H_{\mathrm{sc}}^{(\mathrm{I})}+\Sigma_{\mathrm{I}}^r(E).
		\label{eq:HeffI}
	\end{equation}
	
	Equations~(\ref{eq:sigma11I})--(\ref{eq:sigma21I}) separate the roles of the three ingredients that control Model~I. The scalar part $g_0(E)$ describes local return processes through the SSH waveguide, whereas $g_x(E)$ measures how the excitation leaves one sublattice and returns through the other. The phases $\phi_1$ and $\phi_2$ do not enter as simple overall shifts; they weight the different return paths and therefore decide how the two qubits interfere at the contact cell. The signs and magnitudes of $\gamma_j$ then determine whether the corresponding dressed resonance acts predominantly as a lossy channel or as an amplifying one.
	
	\subsection{Contact-channel decomposition and the bright/dark scattering mechanism}
	
	The bright/dark structure becomes transparent once the contact cell itself is diagonalized in the symmetric and antisymmetric combinations
	\begin{equation}
		|c_+\rangle=\frac{|A\rangle+|B\rangle}{\sqrt2},\qquad
		|c_-\rangle=\frac{|A\rangle-|B\rangle}{\sqrt2}.
		\label{eq:contact_pm}
	\end{equation}
	Introducing
	\begin{equation}
		U=\frac{1}{\sqrt2}
		\begin{pmatrix}
			1 & 1\\
			1 & -1
		\end{pmatrix},
	\end{equation}
	one obtains
	\begin{align}
		\widetilde g_{00}^r(E)=U^{\dagger}g_{00}^r(E)U&=
		\begin{pmatrix}
			g_+(E) & 0\\
			0 & g_-(E)
		\end{pmatrix},
		\nonumber\\
		g_{\pm}(E)&=g_0(E)\pm g_x(E).
		\label{eq:gpm}
	\end{align}
	The SSH environment therefore resolves the local contact cell into two inequivalent channels. In a uniform waveguide one would have $g_x(E)=0$, so the symmetric and antisymmetric contact combinations would be dressed identically. In the SSH lattice, by contrast, $g_x(E)\neq0$ and the two combinations acquire different dispersive shifts and different radiative widths.
	
	Now specialize to the symmetric interferometer,
	\begin{align}
		J_{1A}&=J_{1B}=J_{2A}=J_{2B}\equiv J,\nonumber\\
		\phi_1&=\phi_2\equiv \phi,~~~
		\tilde\omega_1=\tilde\omega_2\equiv \tilde\omega_q .
		\label{eq:symmetricI}
	\end{align}
	In the qubit molecular basis
	\begin{equation}
		|q_+\rangle=\frac{|e_1\rangle+|e_2\rangle}{\sqrt2},\qquad
		|q_-\rangle=\frac{|e_1\rangle-|e_2\rangle}{\sqrt2},
		\label{eq:brightdark}
	\end{equation}
	the coupling matrix diagonalizes simultaneously with the contact basis:
	\begin{equation}
		\widetilde W_{\mathrm I}=U^{\dagger}W_{\mathrm I}U=
		\ee^{\ii\phi/2}
		\begin{pmatrix}
			2J\cos(\phi/2) & 0\\
			0 & 2\ii J\sin(\phi/2)
		\end{pmatrix}.
		\label{eq:WtildeI}
	\end{equation}
	Therefore the two molecular qubit modes do not couple to the SSH waveguide in the same way. The symmetric molecular mode $|q_+\rangle$ couples exclusively to the symmetric contact channel $|c_+\rangle$ with strength
	\begin{equation}
		\lambda_+(\phi)=2J\ee^{\ii\phi/2}\cos\frac{\phi}{2},
	\end{equation}
	whereas the antisymmetric molecular mode $|q_-\rangle$ couples only to the antisymmetric contact channel $|c_-\rangle$ with strength
	\begin{equation}
		\lambda_-(\phi)=2\ii J\ee^{\ii\phi/2}\sin\frac{\phi}{2}.
	\end{equation}
	The corresponding self-energies become
	\begin{equation}
		\Sigma_+^{(\mathrm{I})}(E)=|\lambda_+(\phi)|^2\,g_+(E)
		=4J^2\cos^2\!\frac{\phi}{2}\,[g_0(E)+g_x(E)],
		\label{eq:sigmaplus}
	\end{equation}
	\begin{equation}
		\Sigma_-^{(\mathrm{I})}(E)=|\lambda_-(\phi)|^2\,g_-(E)
		=4J^2\sin^2\!\frac{\phi}{2}\,[g_0(E)-g_x(E)].
		\label{eq:sigmaminus}
	\end{equation}
	This is the microscopic origin of the bright/dark terminology. At $\phi=0$, $\lambda_-(0)=0$, so the antisymmetric molecular mode does not couple to the continuum at all. At $\phi=\pi$, $\lambda_+(\pi)=0$, and the roles are reversed. Away from these symmetry points both channels are visible, but they are still inequivalent because the $+$ branch samples $g_+(E)$ and the $-$ branch samples $g_-(E)$. The distinction is therefore not only kinematic; it is encoded in the way the SSH environment dresses each branch.
	
	A useful physical point is that the bright and dark molecular modes are not the asymptotic channels of the problem. The incoming object is the SSH Bloch spinor $u_s(k)$ evaluated on the contact cell. Its local weight must first be resolved into the two contact combinations,
	\begin{equation}
		u_s(k)=\alpha_+(k)\,|c_+\rangle+\alpha_-(k)\,|c_-\rangle,
		\label{eq:spinor_decomp}
	\end{equation}
	with weights
	\begin{align}
		\alpha_+(k)&=\langle c_+|u_s(k)\rangle=\frac{1+s\ee^{\ii\theta_k}}{2},
		\nonumber\\
		\alpha_-(k)&=\langle c_-|u_s(k)\rangle=\frac{1-s\ee^{\ii\theta_k}}{2}.
		\label{eq:alpha_pm}
	\end{align}
	The scattering process is therefore two-step. The incident Bloch wave first sets the amplitudes entering $|c_+\rangle$ and $|c_-\rangle$. Those two contact amplitudes then couple independently to $|q_+\rangle$ and $|q_-\rangle$ through $\lambda_+(\phi)$ and $\lambda_-(\phi)$. A compact measure of this channel participation is
	\begin{equation}
		\mathcal P_{\pm}(k,\phi)=|\alpha_{\pm}(k)|^2\,|\lambda_{\pm}(\phi)|^2,
		\label{eq:participation_pm}
	\end{equation}
	which makes the selection rule transparent: a molecular branch is invisible either because the interferometer decouples it from the continuum, $|\lambda_{\pm}|=0$, or because the incoming Bloch spinor carries essentially no weight in the corresponding contact channel, $|\alpha_{\pm}|=0$. In the upper SSH band used below, $|\alpha_+(k)|^2$ remains close to unity whereas $|\alpha_-(k)|^2$ is much smaller. The narrow branch is therefore weak not only because of flux-controlled decoupling, but also because the incoming SSH spinor places little weight in the channel that feeds it.
	
	The $T$ matrix itself inherits the same separation. In the transformed basis one finds
	\begin{equation}
		\widetilde T_{\mathrm I}(E)=
		\begin{pmatrix}
			\tau_+(E) & 0\\
			0 & \tau_-(E)
		\end{pmatrix},
		~~
		\tau_{\pm}(E)=\frac{|\lambda_{\pm}(\phi)|^2}{E-\tilde\omega_q-\Sigma_{\pm}^{(\mathrm I)}(E)}.\nonumber
		%\label{eq:TdiagI}
	\end{equation}
	Thus the poles of the symmetric device are simply
	\begin{equation}
		E-\tilde\omega_q-\Sigma_{\pm}^{(\mathrm{I})}(E)=0.
		\label{eq:poleI_symmetric}
	\end{equation}
	Near but not exactly at $\phi=0$ or $\phi=\pi$, one channel becomes quasi-dark and acquires a very narrow radiative width. In a passive device this gives flux-tunable subradiant resonances. In the non-Hermitian circuit studied here the same narrow mode can instead become a narrow absorption resonance or a sensitive amplifying resonance, because even a small gain/loss imbalance acts on a mode whose radiative leakage has already been strongly suppressed by interference.
	
	\begin{figure*}[t]
		\centering\includegraphics[width=1\linewidth]{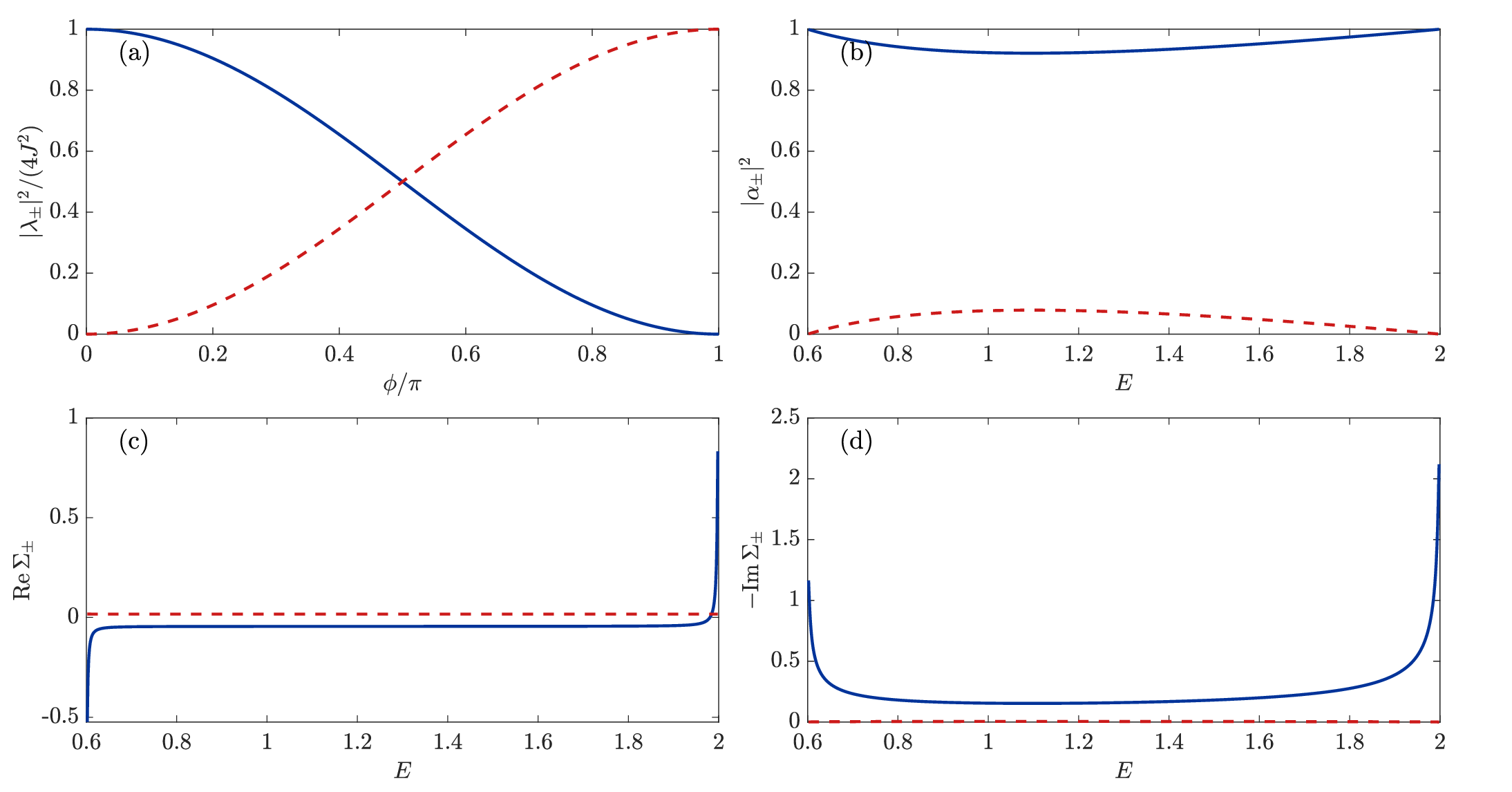} 
		\caption{Channel decomposition of the symmetric Model-I interferometer. The parameters are $t=1$, $\delta=0.30$, $J=0.20$, and $\phi_0=0.35\pi$. Panel (a) shows the flux-controlled molecular couplings $|\lambda_{\pm}(\phi)|^2/(4J^2)$. Panel (b) plots the upper-band contact-channel projections $|\alpha_{\pm}(k)|^2$ and demonstrates that the incoming Bloch spinor remains strongly biased toward the symmetric contact channel. Panels (c) and (d) display the real and imaginary parts of $\Sigma_{\pm}^{(\mathrm I)}(E)$ at $\phi=0.35\pi$. Even when the molecular couplings become comparable, the $+$ branch still carries most of the dispersive pull and most of the radiative width, while the $-$ branch stays almost lossless. The figure therefore shows that the bright/dark distinction is produced jointly by flux control, SSH channel asymmetry, and Bloch-spinor selectivity.}
		\label{fig:modelI_channels}
	\end{figure*}

	Figure~\ref{fig:modelI_channels} makes this channel hierarchy quantitative before one looks at the full scattering spectra. Panel~(a) shows the purely interferometric part: as the flux is increased, the bright $+$ branch is continuously weakened while the $-$ branch is opened. Taken by itself, this panel might suggest that the two branches become comparably important around $\phi\sim 0.35\pi$--$0.50\pi$. Panels~(b)--(d) show why that conclusion would be incomplete in the SSH waveguide.
	
	Panel~(b) gives the second ingredient, namely the contact-spinor selectivity of the incoming upper-band Bloch wave. Over the energy range used below, the incident state projects overwhelmingly onto $|c_+\rangle$ and only weakly onto $|c_-\rangle$. The narrow branch is therefore suppressed already at the injection stage: even when $|\lambda_-(\phi)|$ is no longer very small, the incoming wave does not feed that branch efficiently. Panels~(c) and (d) add the third ingredient, namely the branch-dependent SSH dressing. At the representative flux $\phi=0.35\pi$, the self-energy of the $+$ branch shows both a much larger dispersive variation and a much larger imaginary part than the $-$ branch. In physical terms, the $+$ sector is the branch that the SSH continuum can pull strongly and broaden strongly, whereas the $-$ sector remains only weakly dressed and therefore nearly lossless.
	
	The main lesson of Fig.~\ref{fig:modelI_channels} is that the bright/quasi-dark separation is not controlled by flux alone. It is the combined outcome of three filters acting in sequence: the flux-controlled molecular couplings $\lambda_\pm$, the SSH Bloch-spinor weights $\alpha_\pm$, and the branch-dependent self-energies $\Sigma_\pm^{(\mathrm I)}$. This is why Model~I generically produces one broad dressed feature and one much narrower satellite, rather than two resonances of comparable visibility.
	
	\subsection{Representative scattering spectra and dimerization-sensitive line shapes}
	
	Substituting $T_{\mathrm I}(E)$ into Eqs.~(\ref{eq:tL_main})--(\ref{eq:rR_main}) gives the exact left- and right-incidence amplitudes for the parallel interferometer. In the symmetric basis the left-incidence transmission amplitude can be written as
	\begin{equation}
		t_L(E)=1-\frac{\ii}{v_s(k)}
		\left[
		|\alpha_+(k)|^2\tau_+(E)+|\alpha_-(k)|^2\tau_-(E)
		\right],
		\label{eq:tL_pm}
	\end{equation}
	whereas the reflection amplitude is
	\begin{equation}
		r_L(E)=-\frac{\ii}{v_s(k)}
		\left[
		\beta_+^*(k)\alpha_+(k)\tau_+(E)+
		\beta_-^*(k)\alpha_-(k)\tau_-(E)
		\right],
		\label{eq:rL_pm}
	\end{equation}
	with $\beta_{\pm}(k)=\langle c_{\pm}|u_s(-k)\rangle$. These expressions make the line-shape mechanism explicit. Transmission and reflection are not controlled by a single dressed resonance; they are coherent sums of two molecular contributions, each weighted by how strongly the incoming and outgoing Bloch spinors project onto the corresponding contact channel.
	
	\begin{figure*}[t]
		\centering\includegraphics[width=1\linewidth]{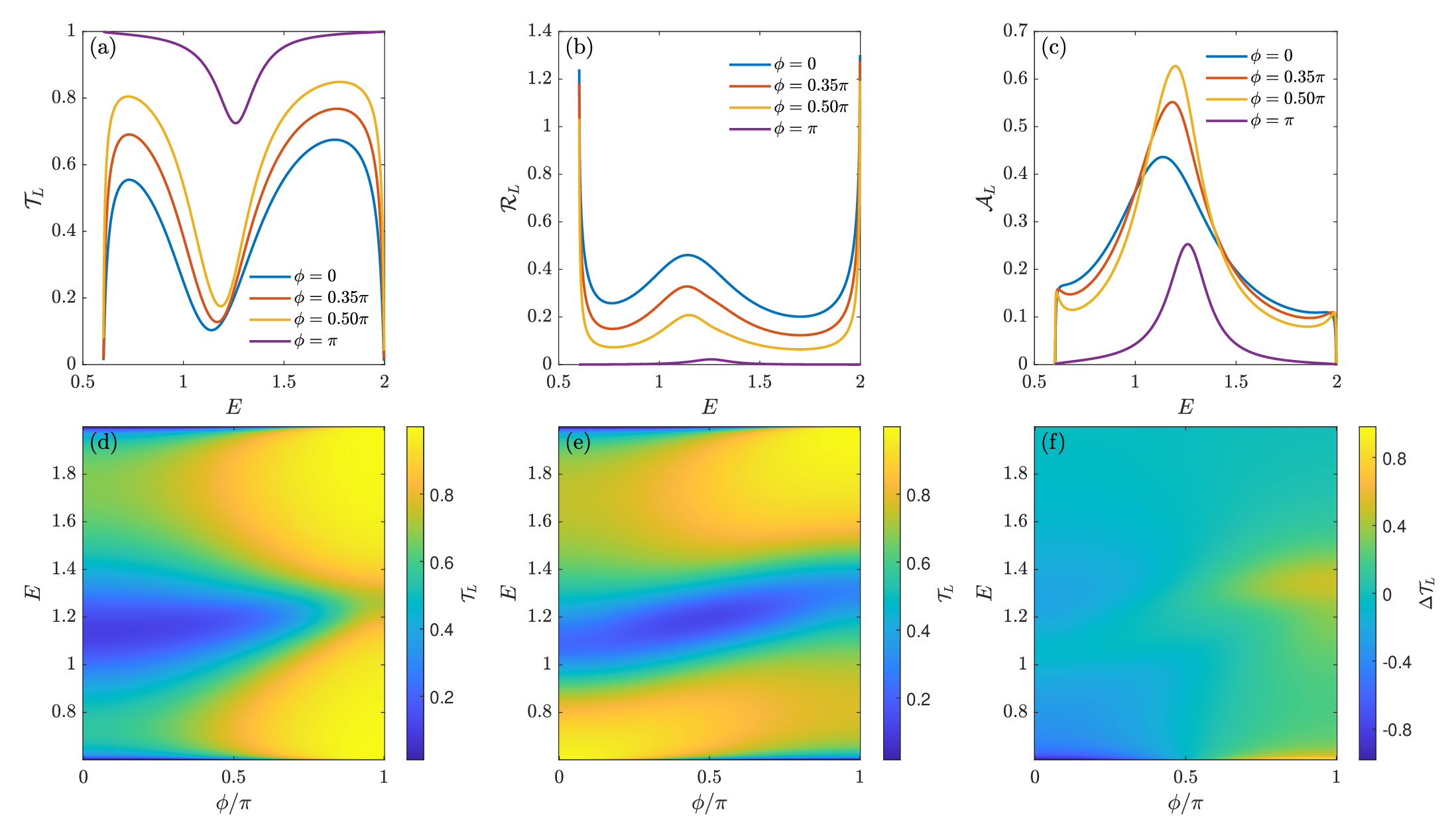}
		\caption{Representative passive scattering spectra of Model I. The parameters are $t=1$, $J=0.20$, $\omega_q=1.20$, and $\gamma_1=\gamma_2=-0.10$. Panels (a)--(c) show the left-incidence transmittance $\mathcal T_L$, reflectance $\mathcal R_L$, and absorptance $\mathcal A_L$ in the upper SSH band for $\phi=0$, $0.35\pi$, $0.50\pi$, and $\pi$, at fixed dimerization $\delta=+0.30$. The line cuts reveal a broad bright-channel stop feature at small flux and a progressively sharper quasi-dark resonance as the weak molecular branch is opened. Panels (d) and (e) display the flux--energy maps of $\mathcal T_L(E,\phi)$ for $\delta=+0.30$ and $\delta=-0.30$, respectively, while panel (f) plots the difference $\Delta\mathcal T_L=\mathcal T_L(\delta=+0.30)-\mathcal T_L(\delta=-0.30)$. The sign-changing lobe pattern around the dressed resonance is a direct dimerization-sensitive interferometric signature of the SSH environment.}
		\label{fig:modelI_spectra}
	\end{figure*}
	
	Figure~\ref{fig:modelI_spectra} shows how this hierarchy appears in passive spectra for the representative point $J=0.20$, $\omega_q=1.20$, and $\gamma_1=\gamma_2=-0.10$. At $\phi=0$ the dark branch is exactly decoupled, so the response is set by one broad bright-channel resonance. Increasing the flux to $0.35\pi$ and $0.50\pi$ opens the weak branch. It does not produce a second broad stop band. Instead, it cuts a narrow interference structure into the bright background and enhances the absorption near the middle of the dressed feature. At $\phi=\pi$ the broad branch is switched off, yet a narrow residual absorption line survives because the incoming Bloch spinor still has a finite projection onto the contact channel that feeds the quasi-dark state.
	
	The same figure also compares the two dimerizations $\delta=\pm0.30$. Reversing the dimerization changes the local SSH Green function and therefore changes how both the broad and narrow branches are dressed. In the flux--energy maps this appears mainly as a shift and bending of the principal transmission minimum near $E\approx1.2$. The difference map $\Delta\mathcal T_L=\mathcal T_L(\delta=+0.30)-\mathcal T_L(\delta=-0.30)$ makes the effect visible as a sign-changing lobe structure around the dressed resonance. In Model~I the topological contrast is therefore clear but not extreme: the SSH dimerization reorganizes the detailed line shape and moves the best operating point, while the overall bright/quasi-dark hierarchy remains intact.
	
\begin{figure*}[t]
		\centering\includegraphics[width=1\linewidth]{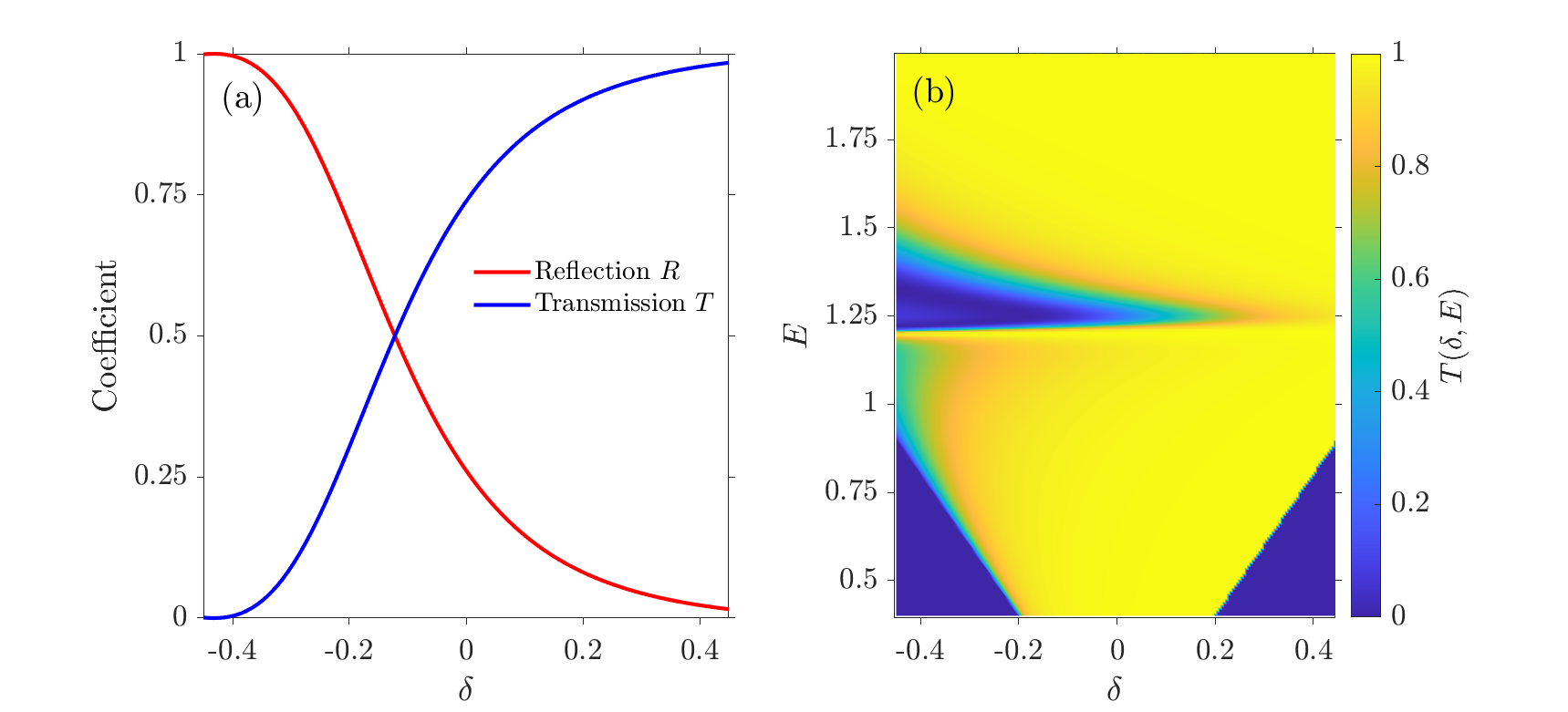}
		\caption{Dimerization-dependent scattering response of Model I. The parameters are $t=1$, $J=0.20$, $\omega_q=1.20$, and $\gamma_1=-\gamma_2=-0.05$. (a) left-incidence transmittance $\mathcal T_L$, reflectance $\mathcal R_L$, as functions of the SSH dimerization parameter $\delta$ at the representative switching energy $E_{\rm sw}\simeq1.32$ and synthetic flux $\phi=\pi$. The transport response evolves continuously with $\delta$, exhibiting a crossover from reflection-dominated to transmission-dominated behavior as the SSH dimerization is varied. (b) density map of the transmission coefficient $T_L(E,\delta)$ in the energy–dimerization plane. }
		\label{fig:T_delta}
	\end{figure*}

	To further clarify the role of the SSH dimerization, Fig.~\ref{fig:T_delta}(a) presents the directly measurable left-incidence transmittance $\mathcal T_L$, reflectance $\mathcal R_L$  as functions of the dimerization parameter $\delta$ at the representative switching energy $E_{\rm sw}\simeq1.32$ and $\phi=\pi$. The transmission and reflection evolve smoothly as the dimerization is varied across the SSH transition point. For negative dimerization, the scattering response is predominantly reflection dominated, whereas positive dimerization favors transmission. The crossover between the two behaviors occurs continuously near $\delta \approx -0.1$, where the transmission and reflection become comparable. 
	The corresponding $E$-$\delta$ map of $\mathcal T_L$ shown in Fig.~\ref{fig:T_delta}(b) demonstrates that this contrast is not confined to the two representative values $\delta=\pm 0.30$ (used in Fig.~\ref{fig:modelI_spectra}) adopted in the previous discussion. Instead, a finite region of the $(E,\delta)$ parameter space exhibits a pronounced dimerization-dependent transport response. At the same time, the map clearly shows that the effect is restricted to specific energy windows associated with the SSH-induced dressing of the local scattering states and does not extend uniformly across the entire propagating band. 
	Importantly, neither dataset displays a quantized jump nor a universal discontinuity at $\delta=0$, demonstrating that the observed transport contrast does not originate from the opening and closing of the SSH bulk gap at $k=\pi$. For the entire range of switching energies considered in this work, including $E_{\rm sw}$, incident photons propagate within the continuous band of the SSH waveguide, so that scattering events are governed by extended Bloch modes instead of gap states. The observed pass/reflect contrast should therefore be understood as a dimerization-sensitive interference effect mediated by the SSH self-energy. While the SSH dimerization strongly reshapes the scattering response of the same local superconducting circuit, the transport coefficients themselves are not topological invariants and consequently do not constitute a protected topological switching effect.

	It is worth noting that, all key parameters in our theoretical models are experimentally accessible in state-of-the-art superconducting circuit QED platforms. The dimerization strength $\delta$ (i.e., $t_1=t(1+\delta), t_2=t(1-\delta)$) can be realized by tunable capacitive or inductive couplers between resonators, which continuously tune intra-cell and inter-cell hopping amplitudes. Synthetic fluxes $\phi_{1, 2}$ are implemented via flux-biased superconducting quantum interference devices or parametric modulators, allowing continuous tuning from $0$ to $\pi$. Balanced gain and loss ($\gamma_1=-\gamma_2$) can be engineered by coupling each qubit to a separate reservoir. Implemented via reservoir engineering and parametric pumping on superconducting qubits and resonators, providing controllable effective gain and dissipative loss to realize balanced non-Hermitian configurations. Thus the parameter values used in our numerical simulations are within current experimental reach, and offering direct theoretical guidance for future prototype realization.
	
	\subsection{Exceptional points, CPA precursors, and zero-pole diagnostics}
	
	Away from the fully symmetric limit the bright and dark sectors hybridize. The effective Hamiltonian then takes the generic two-by-two form
	\begin{equation}
		H_{\mathrm{eff}}^{(\mathrm{I})}(E)=
		\begin{pmatrix}
			h_{11}(E) & h_{12}(E)\\
			h_{21}(E) & h_{22}(E)
		\end{pmatrix},
		\label{eq:HeffI_full}
	\end{equation}
	and a second-order exceptional point satisfies
	\begin{equation}
		\Big[h_{11}(E)-h_{22}(E)\Big]^2+4h_{12}(E)h_{21}(E)=0,
		\label{eq:EP_I}
	\end{equation}
	together with defectiveness of the matrix. In the parallel interferometer this condition is directly accessible because all entries of $H_{\mathrm{eff}}^{(\mathrm{I})}(E)$ can be tuned experimentally by changing the qubit frequencies, the gain/loss imbalance, and the synthetic flux.
	
\begin{figure}[t]
		\centering\includegraphics[width=1\linewidth]{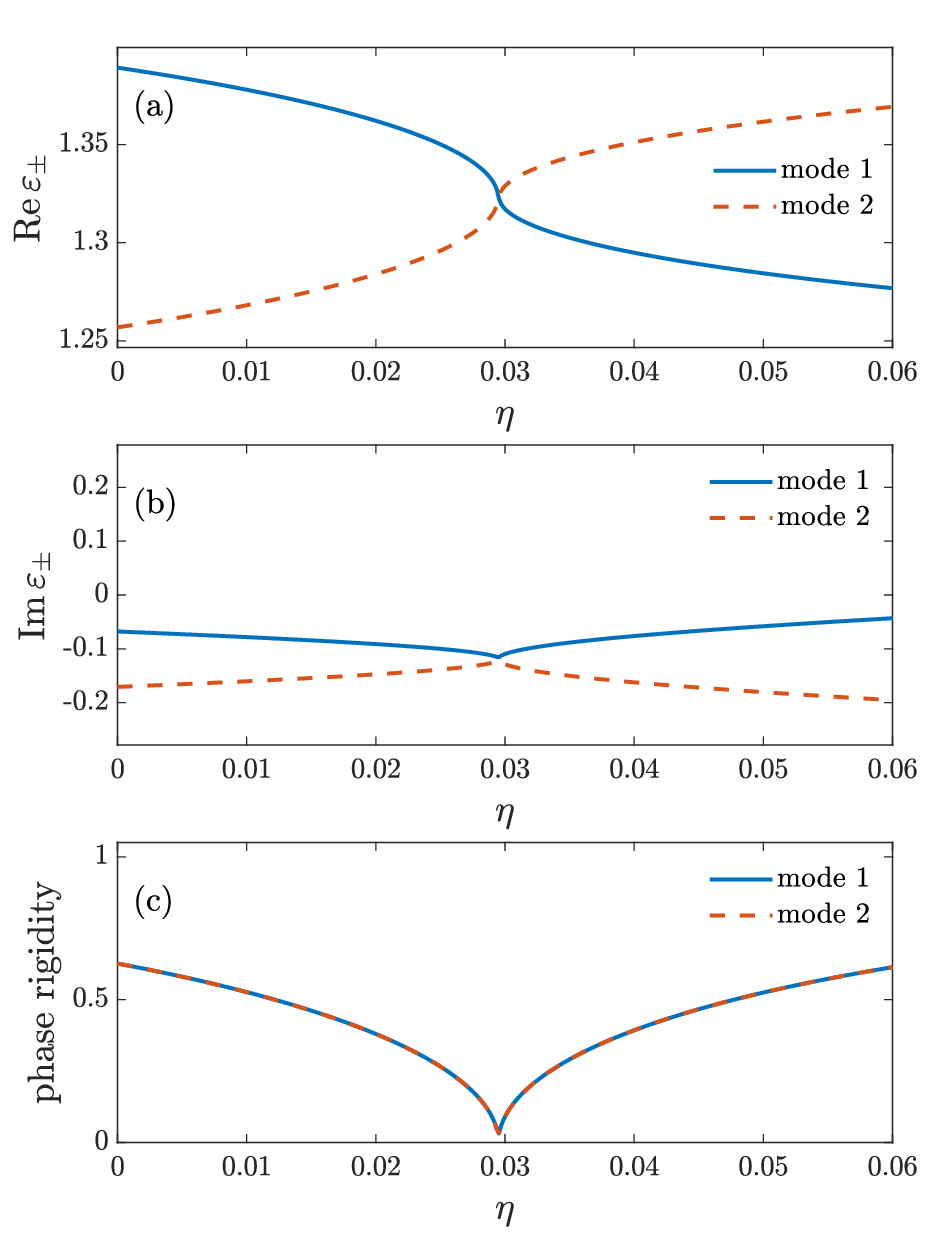}
		\caption{Near-EP diagnostics for Model I. The figure leaves the perfectly symmetric bright/dark limit and uses the representative asymmetric parameter set $t=1$, $\delta=0.30$, $E_*=0.75$, $(J_{1A},J_{1B},J_{2A},J_{2B})=(0.185,0.184,0.250,0.254)$, $(\phi_1,\phi_2)=(0.150\pi,0.413\pi)$, and $(\omega_1,\omega_2)=(1.242,1.465)$. The balanced gain/loss strengths are swept as $\gamma_1=+\eta$ and $\gamma_2=-\eta$. Panels (a) and (b) show the real and imaginary parts of the two eigenvalues of $H_{\mathrm{eff}}^{(\mathrm I)}(E_*)$, while panel (c) displays the biorthogonal phase rigidities. The simultaneous spectral approach and pronounced rigidity suppression identify a controlled near-EP regime generated by bright--dark hybridization once the protective symmetry is relaxed.}
		\label{fig:modelI_EP}
	\end{figure}

	A genuine second-order EP cannot occur in the perfectly symmetric bright/dark limit, because the two molecular channels are then exactly decoupled. An EP requires both non-Hermiticity and channel mixing, so one must first relax the symmetry that protects the decomposition. A numerical scan over couplings, phase distributions, qubit detunings, and balanced gain/loss strengths identifies the representative working point
	\begin{align}
		(J_{1A},J_{1B},J_{2A},J_{2B})&=(0.185,0.184,0.250,0.254),\nonumber\\
		(\phi_1,\phi_2)&=(0.150\pi,0.413\pi),\nonumber\\
		(\omega_1,\omega_2)&=(1.242,1.465). \nonumber
	\end{align}
	produces a robust near-EP regime already for moderate non-Hermitian rates. Figure~\ref{fig:modelI_EP} sweeps $\gamma_1=+\eta$ and $\gamma_2=-\eta$ at this working point. The real parts of the two eigenvalues move toward one another, the imaginary parts nearly coalesce, and the biorthogonal phase rigidity drops far below unity. In the plotted one-dimensional cut the strongest suppression occurs in a narrow low-$\eta$ corridor, while a broader hybridization region persists at intermediate $\eta$. This is the expected signature of a cut through an EP neighborhood at fixed real probe energy. In the full multidimensional parameter space the exact algebraic EP is generally located slightly away from the one-dimensional scan shown here, so the figure should be read as a controlled passage through the EP corridor rather than as evidence of exact coalescence on the plotted axis.
	
	The same two-port formulation also gives direct access to scattering zeros and poles. In the present model a real-energy CPA point requires $\det S(E)=0$, while a lasing threshold corresponds to a pole of $S(E)$ reaching the real axis. In practice the numerically most stable diagnostics are the smallest and largest singular values of $S(E)$: the former tends to zero at CPA and the latter diverges near a lasing pole. Because both are inherited from the same effective Hamiltonian, they can approach one another in a narrow region of parameter space even before a true CPA-laser point is reached.
	
\begin{figure*}[t]
		\centering\includegraphics[width=1\linewidth]{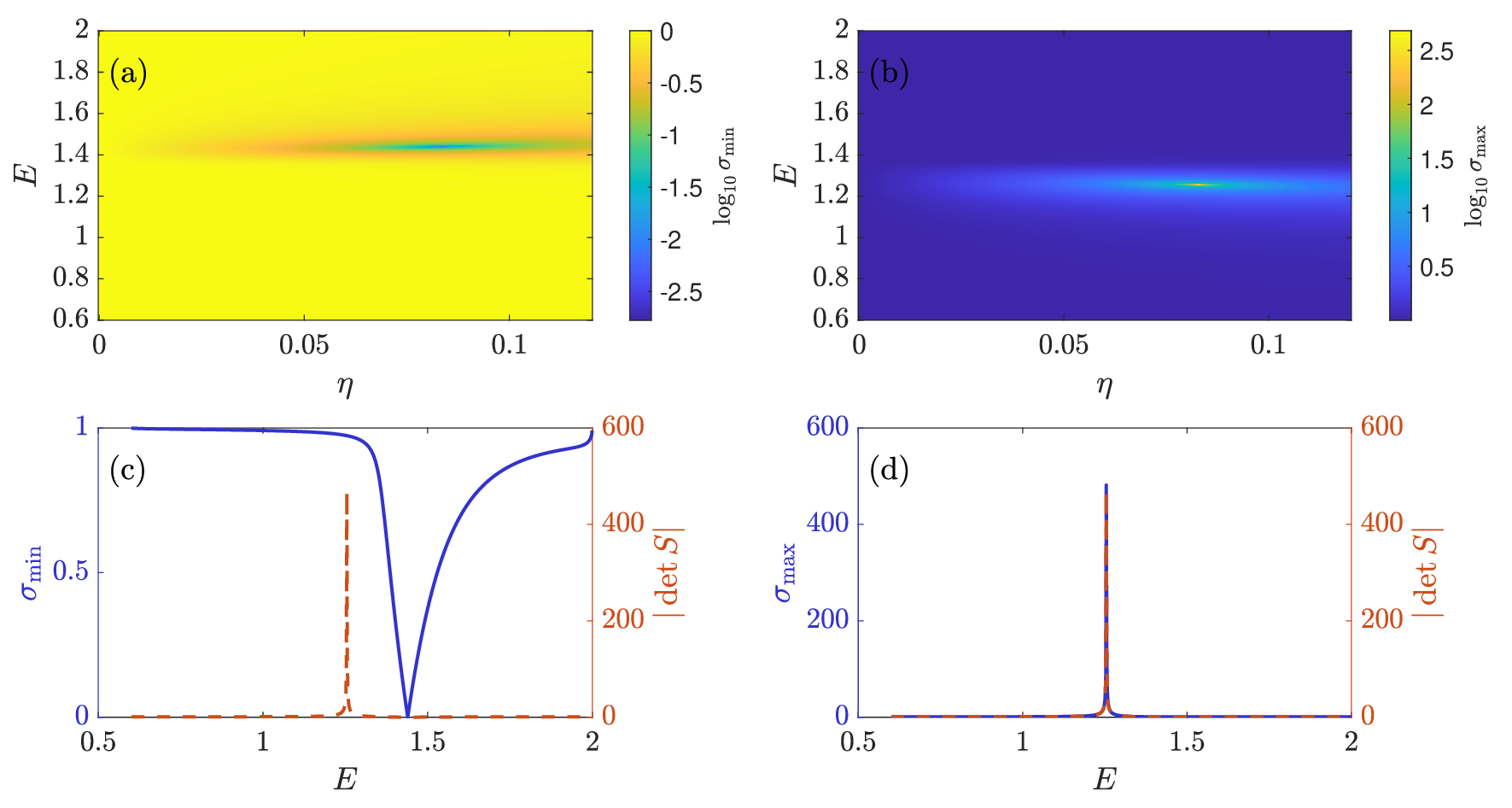}
		\caption{Zero-pole diagnostics of the Model-I scattering matrix. The parameters are the same asymmetric ones used in Fig.~\ref{fig:modelI_EP}, and the balanced gain/loss strengths are swept as $\gamma_1=+\eta$ and $\gamma_2=-\eta$ over a broad interval. Panel (a) shows $\log_{10}\sigma_{\min}[S(E)]$ in the $(E,\eta)$ plane, while panel (b) shows $\log_{10}\sigma_{\max}[S(E)]$. Panels (c) and (d) display the automatically selected cuts through the global minimum of $\sigma_{\min}$ and the global maximum of $\sigma_{\max}$ on the scan grid, together with $|\det S(E)|$. The map demonstrates that the strongest near-zero singular-value response is carried by the narrow quasi-dark sector, whereas the largest amplification response is carried by the bright-hybridized sector.}
		\label{fig:modelI_zeropolepole}
	\end{figure*}

	Figure~\ref{fig:modelI_zeropolepole} maps these two singular values at the same asymmetric working point used in the near-EP scan, but now over a wider gain/loss interval. This broader scan is natural for superconducting-circuit implementations because engineered dissipation or gain can be varied over a large dynamic range by reservoir engineering, and from a theoretical perspective it would be unnecessarily restrictive to keep $\eta$ infinitesimal. The map shows a clear branch separation. 
	Notably, the SSH continuum self-energy is evaluated using the closed-form retarded Green functions of the SSH lattice rather than direct numerical quadrature of the continuum integral.
	The deepest minimum of the smallest singular value sits on the narrow quasi-dark side of the dressed spectrum, whereas the largest value of $\sigma_{\max}$ is shifted toward the more strongly hybridized bright side. 
	Scattering zeros and pole-dominated responses are therefore routed to different regions of the molecular spectrum by the same channel hierarchy that shapes the passive line cuts. From a device perspective this is useful: absorption-like and amplification-like operating points can be targeted separately by deciding whether the circuit is biased toward the narrow branch or the broad one.
	
	Model~I is therefore governed by a simple but robust structure. In the symmetric limit the interferometer separates into a broad bright molecular channel and a narrow dark or quasi-dark one; their visibility is fixed jointly by the synthetic flux and by the Bloch-spinor content of the SSH contact state. Weak symmetry breaking mixes these two channels, generates a near-EP corridor, and separates near-zero and pole-dominated responses across different parts of the dressed spectrum.

	\section{Model II: two qubits linked by an auxiliary internal mode}
	\label{sec:modelII}
	
	\subsection{Microscopic Hamiltonian and exact reduction}
	
	The second circuit geometry, shown in Fig.~\ref{fig:models_schematic}(b), extends Model I by introducing an auxiliary internal mode $c$ that couples to both qubits but not directly to the SSH waveguide. Physically, $c$ may represent an auxiliary resonator, a coupler mode, or an intentionally inserted internal cavity inside the superconducting scattering center. In the single-excitation basis $X=(\ket{e_1},\ket{c},\ket{e_2})^T$, the internal Hamiltonian reads
	\begin{equation}
		H_{\mathrm{sc}}^{(\mathrm{II})}=
		\begin{pmatrix}
			\tilde\omega_1 & J_{1c} & 0\\
			J_{1c} & \tilde\omega_c & J_{2c}\\
			0 & J_{2c} & \tilde\omega_2
		\end{pmatrix},
		\qquad
		\tilde\omega_\mu=\omega_\mu+\ii\gamma_\mu,
		\label{eq:HscII}
	\end{equation}
	where $\mu=1,2,c$. The coupling matrix to the SSH contact cell is
	\begin{equation}
		W_{\mathrm{II}}=
		\begin{pmatrix}
			J_{1A}\ee^{\ii\phi_1} & 0 & J_{2A}\\
			J_{1B} & 0 & J_{2B}\ee^{\ii\phi_2}
		\end{pmatrix},
		\label{eq:WII}
	\end{equation}
	so that the mediator remains internal to the scatterer and only the two qubits talk directly to the SSH waveguide.
	
	A central advantage of the Green-function formulation is that the auxiliary mode can be eliminated analytically before coupling the resulting object to the SSH continuum. Partitioning the internal basis into the qubit sector $(e_1,e_2)$ and the auxiliary sector $c$, and taking the Schur complement, yields the exact reduced qubit Hamiltonian
	\begin{equation}
		H_{q,\mathrm{red}}^{(\mathrm{II})}(E)=
		\begin{pmatrix}
			\tilde\omega_1+\dfrac{J_{1c}^2}{E-\tilde\omega_c} &
			\dfrac{J_{1c}J_{2c}}{E-\tilde\omega_c}\\[1.0em]
			\dfrac{J_{1c}J_{2c}}{E-\tilde\omega_c} &
			\tilde\omega_2+\dfrac{J_{2c}^2}{E-\tilde\omega_c}
		\end{pmatrix}.
		\label{eq:HredII}
	\end{equation}
	Equation~(\ref{eq:HredII}) shows directly what the internal mediator does. It contributes both an energy-dependent diagonal dressing and an energy-dependent complex interqubit coupling,
	\begin{equation}
		J_{12}^{\mathrm{eff}}(E)=\frac{J_{1c}J_{2c}}{E-\tilde\omega_c},
		\label{eq:JeffII}
	\end{equation}
	whose real part governs coherent exchange while the imaginary part governs mediator-induced dissipative hybridization. When $\gamma_c\neq0$, the auxiliary mode is therefore a resonant non-Hermitian element that changes both the positions and the widths of the dressed qubit resonances.
	
	\subsection{Mediator-dark hierarchy in the symmetric limit}
	
	The most revealing analytical limit is the symmetric internal configuration
	\begin{equation}
		\tilde\omega_1=\tilde\omega_2\equiv \tilde\omega_q,
		\qquad
		J_{1c}=J_{2c}\equiv J_c.
		\label{eq:symm_internal_cond}
	\end{equation}
	In that case it is natural to define molecular qubit combinations
	\begin{equation}
		\ket{q_\pm}=\frac{\ket{e_1}\pm\ket{e_2}}{\sqrt2}.
		\label{eq:qpm_II}
	\end{equation}
	In the basis $(q_+,c,q_-)$, the isolated internal Hamiltonian becomes
	\begin{equation}
		H_{\mathrm{sc}}^{(\mathrm{II})}\rightarrow
		\begin{pmatrix}
			\tilde\omega_q & \sqrt2\,J_c & 0\\
			\sqrt2\,J_c & \tilde\omega_c & 0\\
			0 & 0 & \tilde\omega_q
		\end{pmatrix}.
		\label{eq:HII_sym_basis}
	\end{equation}
	The qubit-difference mode $q_-$ is therefore exactly dark with respect to the auxiliary mediator: it does not hybridize with $c$, whereas the $q_+$ branch and the auxiliary mode form a polaritonic doublet. After integrating out $c$, this structure reappears as
	\begin{equation}
		H_{q,\mathrm{red}}^{(\mathrm{II})}(E)\rightarrow
		\begin{pmatrix}
			\tilde\omega_q+\dfrac{2J_c^2}{E-\tilde\omega_c} & 0\\[1.0em]
			0 & \tilde\omega_q
		\end{pmatrix}_{(q_+,q_-)},
		\label{eq:HredII_sym_basis}
	\end{equation}
	showing that the mediator dresses only the $q_+$ branch, while $q_-$ remains undressed at the level of the internal circuit.
	
	The significance of this result becomes clear after the SSH waveguide is attached. The waveguide still distinguishes the two contact channels $c_\pm$ introduced in Model~I, and the synthetic flux controls how strongly $q_+$ and $q_-$ couple to them. Model~II therefore contains two nested hierarchies: an internal mediator-bright/mediator-dark structure and, on top of it, the SSH contact-channel structure. The resulting scattering problem is genuinely two-stage: a broad mediator-bright polaritonic sector coexists with a mediator-dark qubit branch whose visibility is then further filtered by the SSH contact channels.
	
	\begin{figure*}[t]
		\centering\includegraphics[width=1\linewidth]{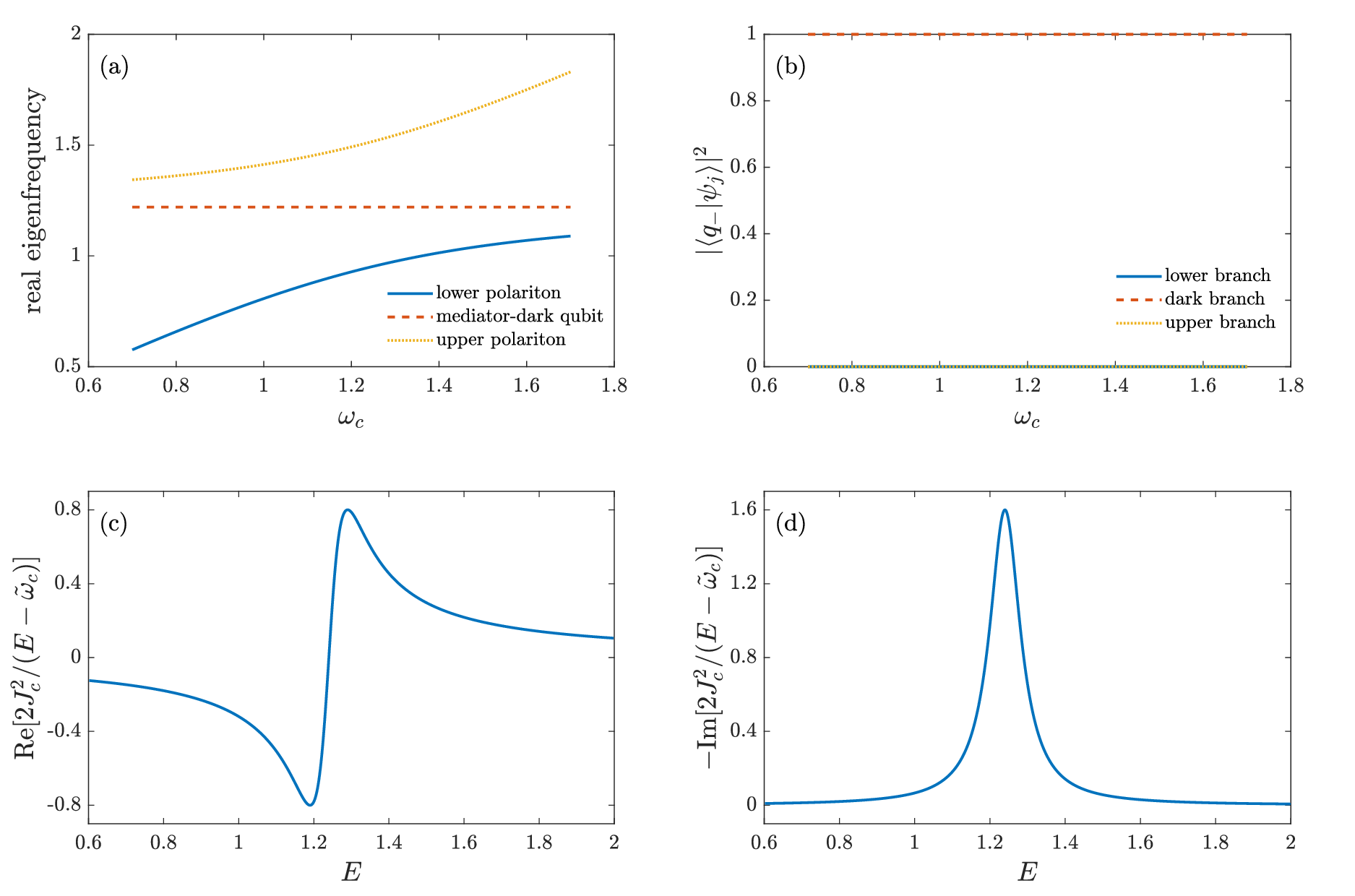}
		\caption{Internal hierarchy of Model II. Panels (a) and (b) use the isolated scatterer in the symmetric limit $\omega_1=\omega_2=\omega_q=1.22$ and $J_{1c}=J_{2c}=J_c=0.20$, while the auxiliary frequency $\omega_c$ is swept. Panel (a) shows the real parts of the three isolated eigenfrequencies. Two branches undergo the expected avoided crossing and form the mediator-bright polaritonic doublet, whereas the third branch stays pinned to $\omega_q$ and is exactly mediator-dark. Panel (b) plots the weight of the qubit-difference state $\ket{q_-}$ on each branch and confirms that one branch remains purely dark throughout the sweep. Panels (c) and (d) evaluate the mediator-induced self-energy $2J_c^2/(E-\tilde\omega_c)$ at fixed $\omega_c=1.24$ and $\gamma_c=-0.05$, displaying its real and imaginary parts across the upper SSH band. The dispersive variation and linewidth enhancement near $E\simeq\omega_c$ are the microscopic origin of the Fano-like and transparency-like structures observed in the full scattering spectra.}
		\label{fig:modelII_hierarchy}
	\end{figure*}
	
	Figure~\ref{fig:modelII_hierarchy} verifies this hierarchy directly. In the isolated scatterer one branch stays nearly pinned to the bare qubit frequency because it is dark to the mediator, while the other two branches undergo the expected avoided crossing and form a polaritonic pair. Panels (c) and (d) show the corresponding resonant self-energy seen by the qubit sector. As the probe energy approaches the auxiliary resonance, both the dispersive shift and the induced linewidth vary rapidly across the upper SSH band. This is the first qualitative difference from Model~I. There the only dressing comes from the SSH waveguide. Here the qubits carry, in addition, an internal resonant self-energy generated by the mediator itself. Depending on flux and detuning, the mediator-induced dressing and the SSH self-energy can reinforce one another or act in opposition. That competition is what produces the broader range of line shapes discussed below.
	
	\subsection{SSH dressing and effective Hamiltonian}
	
	Because only the two qubit states couple directly to the SSH contact cell, the SSH self-energy retains the same formal structure as in Model I but now acts on the reduced qubit subspace,
	\begin{equation}
		\Sigma_{\mathrm{II}}^r(E)=W_q^{\dagger}g_{00}^r(E)W_q,
		\qquad
		W_q=
		\begin{pmatrix}
			J_{1A}\ee^{\ii\phi_1} & J_{2A}\\
			J_{1B} & J_{2B}\ee^{\ii\phi_2}
		\end{pmatrix}.
		\label{eq:SigmaII}
	\end{equation}
	The physically relevant finite-dimensional problem is therefore
	\begin{equation}
		H_{\mathrm{eff}}^{(\mathrm{II})}(E)=H_{q,\mathrm{red}}^{(\mathrm{II})}(E)+\Sigma_{\mathrm{II}}^r(E),
		\label{eq:HeffII}
	\end{equation}
	with the corresponding scattering kernel
	\begin{equation}
		T_{\mathrm{II}}(E)=W_q\Big[E-H_{\mathrm{eff}}^{(\mathrm{II})}(E)\Big]^{-1}W_q^{\dagger}.
		\label{eq:TII}
	\end{equation}
	Equations~(\ref{eq:HredII})--(\ref{eq:TII}) summarize Model~II. The SSH waveguide contributes the matrix self-energy $\Sigma_{\mathrm{II}}^r(E)$, whose off-diagonal part inherits the dimerization-sensitive function $g_x(E)$. The auxiliary mode contributes the resonant factor $J_{12}^{\mathrm{eff}}(E)$ together with the associated diagonal shifts. These two ingredients enter differently and can therefore be tuned with a useful degree of independence.

	\subsection{Passive spectra: mediator-induced Fano, transparency reopening, and dimerization-sensitive transmission/reflection switching}
	
	We now switch to a passive and symmetric operating point chosen specifically to make the mediator-induced line-shape physics visually unambiguous:
	\begin{align}
		J_{1A}&=J_{1B}=J_{2A}=J_{2B}=0.18,~~~J_{1c}=J_{2c}=0.20,\nonumber\\
		\gamma_1&=\gamma_2=0,~~~\gamma_c=-0.05,\nonumber\\
		\omega_1&=\omega_2=1.22,~~~\omega_c=1.24,\nonumber\\
		t&=1,\qquad \delta=\pm0.30,
	\end{align}
	%\[
	%t=1,\qquad \delta=\pm0.30,\qquad
	%J_{1A}=J_{1B}=J_{2A}=J_{2B}=0.18,
	%\]
	%\[
	%\omega_1=\omega_2=1.22,\qquad
	%\omega_c=1.24,\qquad
	%J_{1c}=J_{2c}=0.20,
	%\]
	%\[
	%\gamma_1=\gamma_2=0,\qquad
	%\gamma_c=-0.05,
	%\]
	with a common flux choice $\phi_1=\phi_2\equiv\phi$. This choice keeps the qubits nearly lossless while retaining a lossy mediator. As a result, the interference between the mediator-bright polaritonic background and the mediator-dark branch becomes especially clear. At this passive operating point three effects can be isolated cleanly: mediator-induced Fano reshaping, EIT-like transparency reopening, and a dimerization-sensitive pass/reflect contrast.
	
	\begin{figure*}[t]
		\centering\includegraphics[width=1\linewidth]{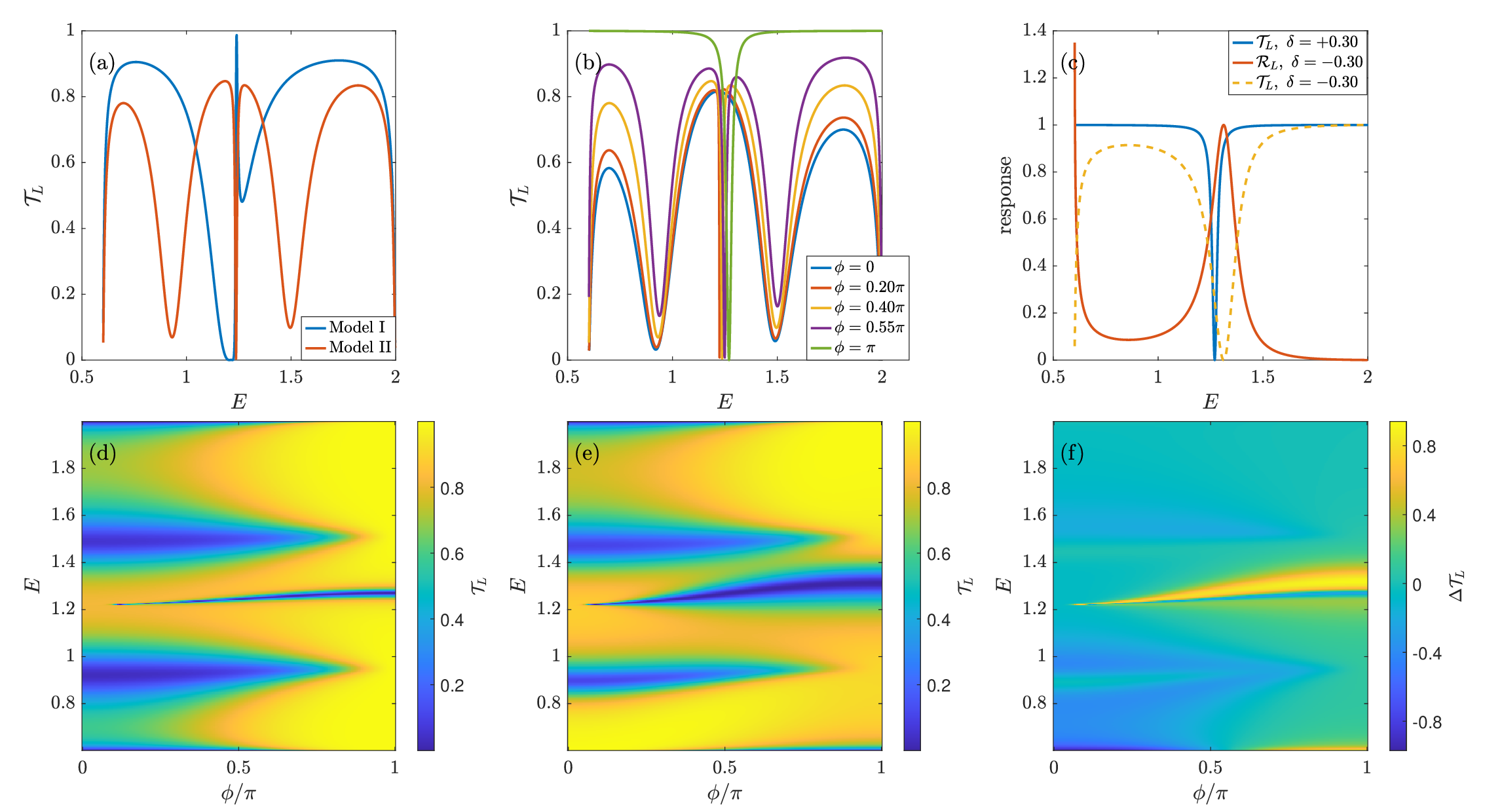}
		\caption{Passive scattering spectra of Model II at a representative passive working point. The parameters are $t=1$, $\omega_1=\omega_2=1.22$, $\omega_c=1.24$, $J_{1A}=J_{1B}=J_{2A}=J_{2B}=0.18$, $J_{1c}=J_{2c}=0.20$, $\gamma_1=\gamma_2=0$, and $\gamma_c=-0.05$. Panel (a) compares the left-incidence transmittance of Model I and Model II at $\phi_*=0.40\pi$ and $\delta=+0.30$; the mediator splits the broad Model-I stop feature into a dressed double-notch structure with a clear transparency reopening between the two mediator-shaped branches. Panel (b) shows Model-II transmittance cuts in the trivial phase ($\delta=+0.30$) for $\phi=0$, $0.20\pi$, $0.40\pi$, $0.55\pi$, and $\pi$, demonstrating the crossover from a single broad dressed suppression to a Fano-like asymmetric dip and then to an EIT-like transparency reopening. Panel (c) compares the two dimerizations at $\phi=\pi$: the trivial lattice retains a high-transmission branch, whereas the nontrivial lattice suppresses transmission at the same energy and becomes nearly perfectly reflective. Panels (d) and (e) display the flux--energy maps of $\mathcal T_L(E,\phi)$ for $\delta=+0.30$ and $\delta=-0.30$, respectively, while panel (f) plots the dimerization-sensitive difference $\Delta\mathcal T_L=\mathcal T_L(\delta=+0.30)-\mathcal T_L(\delta=-0.30)$.}
		\label{fig:modelII_spectra}
	\end{figure*}
	
	Figure~\ref{fig:modelII_spectra}(a) shows the basic spectral effect of the auxiliary mode. In Model~I the dominant structure at this working point is a single broad suppression window. In Model~II that broad feature is split into two dressed structures: a broad background associated with the mediator-bright polaritonic sector and a much narrower feature carried by the weakly radiating branch. The transmission recovery between them is therefore not an accidental shoulder; it is the window left behind when a narrow channel interferes destructively with a broad one.
	
	The flux evolution in Fig.~\ref{fig:modelII_spectra}(b) makes this mechanism transparent. At small flux the narrow branch is only weakly coupled to the SSH ports, so the spectrum is governed mainly by the broad polaritonic background. As the flux increases, the narrow branch acquires finite visibility and cuts into that background, first producing the asymmetric dip characteristic of Fano interference and then reopening a narrow transmission window inside the broader stop region. In physical terms, the broad branch sets the baseline extinction, while the narrow branch locally cancels it. This is the origin of the EIT-like profile in the present passive microwave setting.
	
	Figure~\ref{fig:modelII_spectra}(c) displays the most useful dimerization-sensitive effect. With our convention, $\delta>0$ is topologically trivial and $\delta<0$ is topologically nontrivial. At $\phi=\pi$ the trivial lattice retains a high-transmission channel near $E\simeq1.32$, whereas the nontrivial lattice converts the same incoming wave almost entirely into reflection. Numerically, at the marked operating point one finds $\mathcal T_L\approx0.93$ for $\delta=+0.30$, but $\mathcal T_L\approx0.01$ and $\mathcal R_L\approx0.99$ for $\delta=-0.30$. The circuit geometry is unchanged; what changes is the SSH dressing of the same local branch. %The pass/reflect contrast is therefore genuinely topology selective rather than simply parameter selective.
	%\textcolor[rgb]{0.7725, 0.3529, 0.0667}{The pass/reflect contrast originates from the SSH-phase-dependent dressing of the same local scattering branch. More precisely, reversing the dimerization changes the sign and magnitude of the sublattice-interference contribution ($g_x(E)$), thereby modifying the effective non-Hermitian Hamiltonian and shifting the interference condition responsible for transport. We emphasize that this contrast is not universal over the entire spectrum. Its visibility depends on the probe energy and synthetic flux and becomes particularly strong in parameter regions where the SSH self-energy varies rapidly. The observed switching should therefore be interpreted as an SSH-phase-sensitive transport effect rather than a universal topological order parameter.}
	
	The flux--energy maps in Fig.~\ref{fig:modelII_spectra}(d)--(f) show that the mediator does not wash out the SSH dependence. Instead it redistributes that dependence over two dressed families: a broad polaritonic valley and a narrow mediator-shaped interference line. The difference map $\Delta\mathcal T_L$ accordingly develops a broad sign-changing region together with a much sharper narrow stripe. The broad structure tracks the polaritonic sector, while the narrow stripe tracks the weakly radiating branch that is responsible for the Fano and EIT-like features.
	
	This parameter regime is especially instructive because it also contains a \emph{double-dark} corridor. In the symmetric limit the state $q_-$ is exactly dark to the internal mediator. Near destructive interferometric fluxes the same branch is also only weakly visible to the SSH ports. The linewidth is then suppressed from both sides: it is protected internally because it does not hybridize with $c$, and it is protected externally because it radiates only weakly into the waveguide. The unusually sharp spectral feature that rides on top of the broad background is the direct signature of this doubly protected branch.
	
	\subsection{Active regime: near-EP physics and EP-assisted zero-pole separation}
	
	To expose the genuinely non-Hermitian advantages of Model~II we next move away from the symmetric passive point and allow mild asymmetry in the qubit--waveguide couplings, qubit frequencies, and synthetic phases. A targeted numerical parameter scan was used to identify a robust and visually clean working point,
	\begin{align}
		&(J_{1A},J_{1B},J_{2A},J_{2B})=(0.215,0.181,0.213,0.163),\nonumber\\
		&(\phi_1,\phi_2)=(0.849\pi,0.450\pi),~~~ (J_{1c},J_{2c})=(0.217,0.201),\nonumber\\
		&(\omega_1,\omega_2,\omega_c)=(1.070,1.095,1.340),\nonumber\\
		&\gamma_c=-0.0476,\qquad
		E_*=1.22884.
	\end{align}
	At this working point we sweep balanced qubit gain/loss,
	\begin{align}
		\gamma_1=+\eta,\qquad \gamma_2=-\eta,
	\end{align}
	while keeping the mediator lossy.
	
	\begin{figure*}[t]
		\centering\includegraphics[width=1\linewidth]{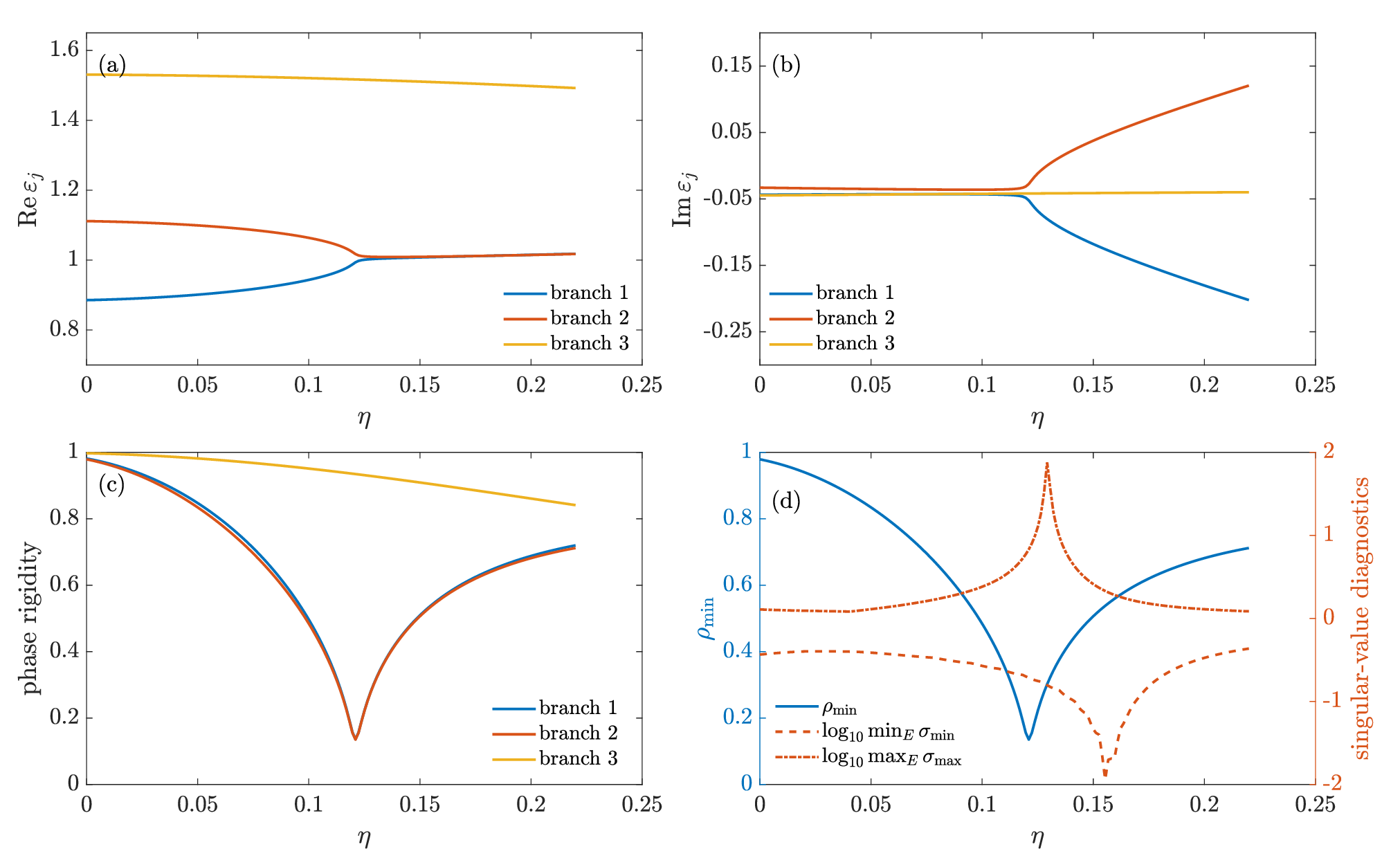}
		\caption{Near-EP diagnostics for Model II together with EP-assisted zero/pole indicators. The asymmetric working point is $t=1$, $\delta=0.30$, $E_*=1.22884$, $(J_{1A},J_{1B},J_{2A},J_{2B})=(0.215,0.181,0.213,0.163)$, $(\phi_1,\phi_2)=(0.849\pi,0.450\pi)$, $(\omega_1,\omega_2,\omega_c)=(1.070,1.095,1.340)$, $(J_{1c},J_{2c})=(0.217,0.201)$, and $\gamma_c=-0.0476$. The balanced qubit gain/loss strengths are swept as $\gamma_1=+\eta$ and $\gamma_2=-\eta$. Panels (a) and (b) show the real and imaginary parts of the three eigenvalues of $H_{\mathrm{eff}}^{(\mathrm{II})}(E_*)$, while panel (c) displays the biorthogonal phase rigidities of the tracked branches. Panel (d) compares three corridor diagnostics extracted from the same $\eta$ scan: the minimum phase rigidity $\rho_{\min}$, the minimum singular value floor $\min_E\sigma_{\min}[S(E)]$, and the largest maximum of the largest singular value $\max_E\sigma_{\max}[S(E)]$. The rigidity collapse around $\eta\simeq0.12$ first sharpens the pole-dominated response and then seeds the deepest near-CPA valley at slightly larger gain/loss bias.}
		\label{fig:modelII_EP}
	\end{figure*}
	
	Figure~\ref{fig:modelII_EP}(a)--(c) shows that the non-Hermitian interaction is concentrated in a two-branch subspace. One branch remains relatively inert over the scan and acts mainly as a spectator. The other two are pushed toward one another by the combined action of mediator-induced coupling and SSH dressing, and their phase rigidities collapse in the same interval of balanced gain/loss. In contrast to Model~I, where the relevant pair comes from leakage between bright and quasi-dark molecular channels, the active doublet here is already embedded in a three-mode internal hierarchy. The mediator selects the relevant pair before the SSH ports read it out.
	
	Panel~(d) identifies the corresponding EP corridor. The minimum phase rigidity occurs near $\eta\simeq0.12$. In the same interval the largest singular value of the scattering matrix rises sharply, signaling the onset of a pole-dominated response. Only at slightly larger bias does the smallest singular value develop its deepest valley. The sequence matters physically. As the system enters the near-EP corridor, non-normal amplification is sharpened first; only after the same dressed doublet is pushed a little further does the destructive interference needed for the strongest near-CPA response emerge on the neighboring branch. The zero-like and pole-like tendencies therefore belong to the same corridor, but not to the same point within it.
	
	\begin{figure*}[t]
		\centering\includegraphics[width=1\linewidth]{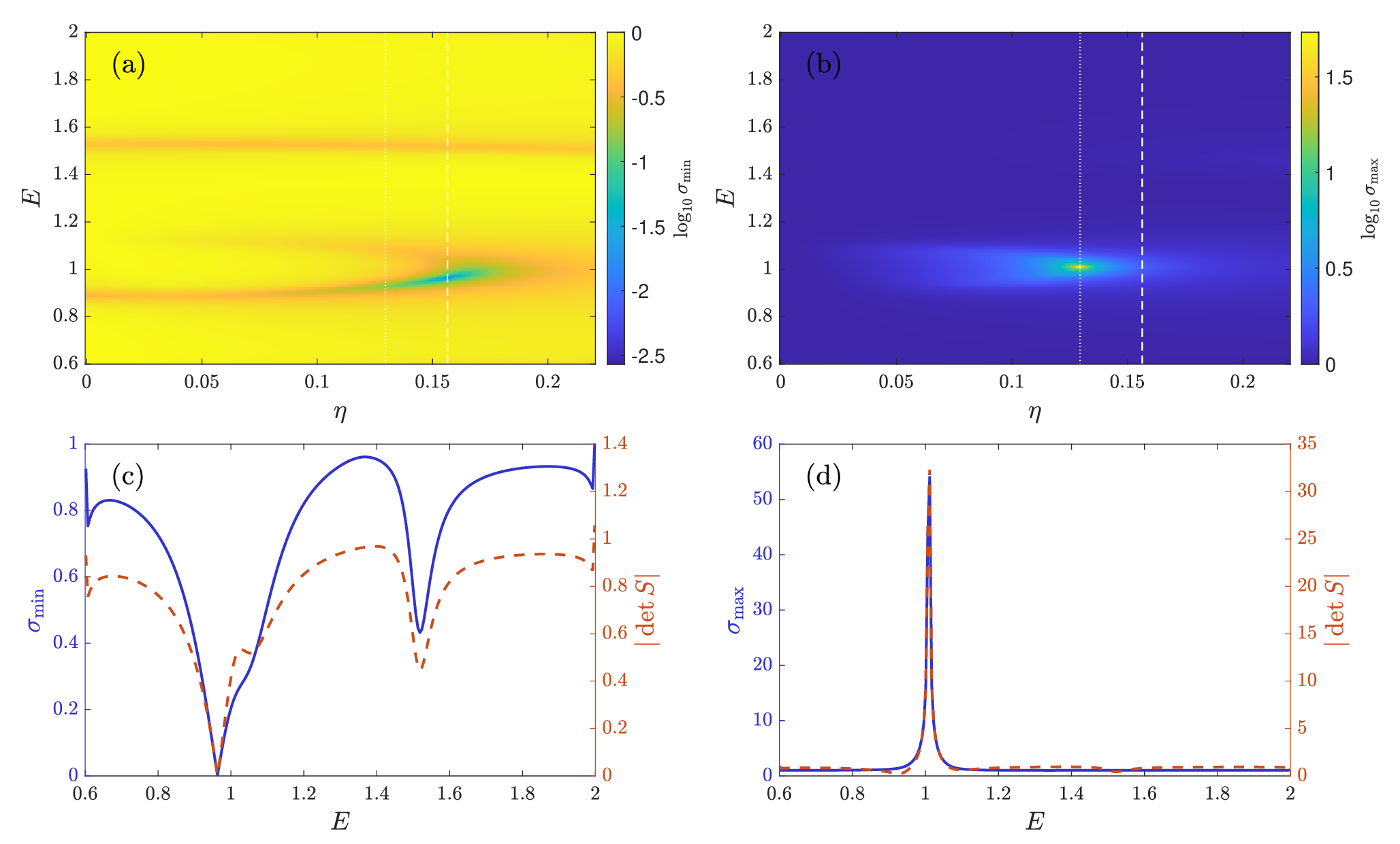}
		\caption{Zero-pole diagnostics of Model II. The same asymmetric working point as in Fig.~\ref{fig:modelII_EP} is used, and the balanced qubit gain/loss strengths are swept over a broad interval. Panel (a) shows $\log_{10}\sigma_{\min}[S(E)]$ and panel (b) shows $\log_{10}\sigma_{\max}[S(E)]$ in the $(E,\eta)$ plane. Panels (c) and (d) display cuts through the strongest near-zero and pole-dominated responses, together with $|\det S(E)|$. The chosen cuts lie inside the same broad near-EP parameter range identified in Fig.~\ref{fig:modelII_EP}(d): the deepest near-CPA valley appears at a slightly larger $\eta$ than the strongest pole-dominated maximum, and the two responses live on neighboring but distinct dressed branches.}
		\label{fig:modelII_zeropolepole}
	\end{figure*}
	
	The maps in Fig.~\ref{fig:modelII_zeropolepole} make this separation explicit. The strongest pole tendency appears first on the upper member of the active doublet, whereas the deepest near-CPA valley develops later on the neighboring lower branch. In other words, the mediator does not merely broaden the design space; it resolves the dressed spectrum finely enough that absorption-like and amplification-like behavior can be routed to adjacent but distinct resonances.
	
	From a device perspective this branch separation is valuable. The hardware does not need to be redesigned in order to move between a threshold-like amplifier and a near-CPA absorber. One instead biases the same circuit to different locations along the same balanced-gain/loss corridor. The earlier ridge emphasizes the pole-dominated branch; the later valley emphasizes the zero-dominated branch. This is precisely the sense in which the auxiliary mode acts as a non-Hermitian branch selector.
	
	The physical picture of Model~II can now be stated compactly. The auxiliary mode first splits the isolated scatterer into a mediator-bright polaritonic sector and a mediator-dark qubit branch. The SSH contact channels then determine how strongly each of those branches appears in transport. In the passive regime this produces Fano reshaping, reopening of a narrow transparency window, and a same-energy pass/reflect switch between the two SSH dimerizations. In the active regime the same hierarchy generates a near-EP corridor and separates zero-dominated and pole-dominated responses across neighboring dressed branches.
	
	\section{From scattering theory to superconducting device functionalities}
	\label{sec:devices}
	
	The analytical structure developed in Sec.~\ref{sec:general} is useful because it turns a spatially extended open-system scattering problem into a finite-dimensional design problem. Once the SSH waveguide is reduced to the two scalar functions $g_0(E)$ and $g_x(E)$, and once the local circuit is specified by $H_{\mathrm{sc}}$ and $W$, all measurable quantities follow from
	\begin{equation}
		H_{\mathrm{eff}}(E)\longrightarrow T(E)\longrightarrow S(E).
	\end{equation}
	This sequence is not merely formal. It makes explicit how local circuit design is converted first into dressed modal structure and then into measurable two-port response. Modal coalescence, scattering zeros, scattering poles, transmission dips, absorption peaks, and amplification ridges are all different manifestations of the same finite-dimensional open-system matrix problem.
	From an implementation perspective, the present framework is directly compatible with currently available superconducting quantum circuits. Tunable couplers allow dynamic control of the SSH dimerization, synthetic gauge phases can be generated by flux-modulated coupling networks, and gain–loss imbalance can be engineered using reservoir-engineering techniques. As a result, the bright–dark interference, mediator-assisted spectral shaping, and near-exceptional-point phenomena predicted here can be explored using existing microwave quantum-simulation architectures.

	\subsection{Exceptional points as dressed and branch-selective design targets}
	
	In the present setting an exceptional point is a property of the dressed matrix $H_{\mathrm{eff}}(E)$, not of the isolated local circuit by itself. This distinction matters in superconducting implementations. A qubit circuit that is far from spectral coalescence in isolation can be pushed into a near-EP regime once the probe energy samples an SSH region where $g_0(E)$ and $g_x(E)$ vary rapidly. Near an SSH band edge, the waveguide provides large dispersive and dissipative corrections, so the degeneracy problem is a joint property of the local circuit and the topological microwave environment.
	
	Model~I and Model~II realize this principle in different ways. In Model~I the relevant pair is created by weakly relaxing the symmetry that otherwise keeps the bright and dark molecular channels independent. The resulting near-EP parameter range is therefore a controlled leakage problem: a nominally dark branch is allowed to mix with a bright one, and the coalescence precursor appears precisely because the protecting symmetry is no longer exact. In Model~II the auxiliary resonator first reorganizes the isolated scatterer into a mediator-bright polaritonic sector and a mediator-dark qubit branch. After SSH dressing is included, one pair inside this three-mode manifold becomes the dominant non-Hermitian doublet while the remaining branch stays comparatively spectator-like. Figures~\ref{fig:modelI_EP} and \ref{fig:modelII_EP} should therefore be read as two distinct EP-design strategies rather than two versions of the same mechanism: Model~I realizes symmetry-relaxed bright--dark hybridization, whereas Model~II realizes branch-selective polaritonic coalescence inside a hierarchical internal spectrum.
	
	A central lesson from Model~II is that the near-EP corridor is directly tied to device response. Figure~\ref{fig:modelII_EP}(d) shows that the same balanced-gain/loss interval that minimizes the phase rigidity first enhances the pole tendency and then deepens the zero-like minimum of the smallest singular value. The strongest non-Hermitian functionality therefore emerges in the same corridor that hosts the strongest spectral non-normality.
	
	\subsection{Scattering zeros, scattering poles, and branch routing in two-port circuits}
	
	The same formalism also clarifies how CPA-like and lasing-like behavior should be diagnosed in a two-port superconducting device. A scattering zero corresponds to a vanishing singular value of $S(E)$ on the real axis, while a pole-dominated response is signaled by a rapidly increasing largest singular value. Since both diagnostics are extracted from the same matrix $S(E)$, they reveal not only whether the device approaches a zero or a pole, but also where in the dressed spectrum that tendency is concentrated.
	
	This branch information is especially useful. In Model~I the strongest near-zero response is tied to the narrow quasi-dark sector, whereas the largest pole-dominated maximum is carried by the brighter hybridized branch. Zeros and pole-dominated responses are therefore not pinned to the same resonance, and absorption-like and amplification-like operation need not occur at the same working point.
	
	Model~II sharpens this separation. Figures~\ref{fig:modelII_EP}(d) and \ref{fig:modelII_zeropolepole} show that the auxiliary mediator reorganizes the dressed resonances so that the strongest pole tendency appears first on one branch, whereas the strongest near-CPA response develops later on the neighboring branch. Strong amplification and strong coherent absorption are thus separated both spectrally and in balanced gain/loss bias.
	
	\subsection{Dimerization-sensitive microwave functionalities}
	
	The numerical results of Secs.~\ref{sec:modelI} and \ref{sec:modelII} point to three concrete microwave functionalities that are especially natural in SSH superconducting circuits.
	
	The first is a dimerization-sensitive reflector or switch. Because the inter-sublattice propagator $g_x(E)$ changes sign and magnitude across the two dimerizations, the same local circuit can produce markedly different line shapes for $\delta>0$ and $\delta<0$. In both models this appears in the sign-changing structure of $\Delta\mathcal T_L$ around the dressed resonances. In Model~II the effect is especially clear: at the passive comparison point highlighted in Fig.~\ref{fig:modelII_spectra}(c), the topologically trivial dimerization ($\delta>0$ in our convention) is nearly transparent, whereas the topologically nontrivial dimerization ($\delta<0$) is nearly perfectly reflective at the same probe energy. The SSH dimerization is therefore not just a background label; it is a genuine topological control knob that can switch the same local superconducting circuit between pass-like and reflect-like operation.
	
	The second is a flux-programmable narrowband filter. In Model~I the relevant object is the quasi-dark molecular branch that appears when one of the flux-controlled couplings $\lambda_\pm(\phi)$ becomes small but not exactly zero. This branch imprints a narrow feature on top of the broad bright-channel background. Model~II upgrades the same idea through its mediator-dark hierarchy. There one branch can be dark with respect to the internal mediator and at the same time only weakly visible to the SSH ports. This doubly protected parameter region produces especially narrow resonant structures and is a natural place to realize notch filters or transparency windows. In this sense, Model~I provides a single-stage quasi-dark filter, whereas Model~II provides a two-stage quasi-dark filter in which internal and external visibility are both suppressed.
	
	The third is branch-selective absorption or amplification. Non-Hermitian parameters do not simply broaden or narrow all resonances together; they redistribute damping and gain among the dressed branches. In Model~I this redistribution already differentiates the quasi-dark and bright-hybridized sectors. In Model~II the mediator makes the redistribution even more selective by first splitting the internal spectrum into a broad polaritonic sector and a much narrower mediator-dark sector, and then allowing the SSH continuum to weight those sectors differently. The result is a useful device principle: the same superconducting circuit can be tuned toward a CPA-like absorber, toward a threshold amplifier, or toward a narrow branch-selective spectral filter, depending on which dressed branch is addressed and how the balanced gain/loss bias is chosen.
	
	More broadly, the theory is constructive as well as descriptive. Once a target function is specified, the formalism indicates which branch should be bright, which one should be protected, where the SSH band structure should be sampled, and whether the objective is better realized by direct interferometric bright/dark engineering (Model~I) or by mediator-assisted hierarchical branch engineering (Model~II).
	\section{Conclusion}
	
	We have developed an exact Green-function formulation for single-microwave-photon scattering in an SSH superconducting waveguide locally coupled to a finite non-Hermitian circuit subsystem. The central result is that the structured SSH environment can be integrated out exactly and recast as an energy-dependent matrix self-energy acting on the finite local device. This reduction turns the full open scattering problem into a finite-dimensional effective non-Hermitian Hamiltonian and places reflection, transmission, absorption, exceptional-point diagnostics, coherent-perfect-absorption conditions, and lasing thresholds within one unified framework.
	
	Within this formulation, Model~I and Model~II reveal two successive layers of environment-shaped scattering physics. In Model~I, a flux-controlled two-qubit interferometric scatterer embedded in the SSH waveguide develops a broad bright branch together with a narrow quasi-dark branch. Their relative weight is determined jointly by the SSH contact structure and the synthetic flux, leading to dimerization-dependent reshaping of the scattering line shape. In particular, the two dimerizations do not simply shift resonances quantitatively; they redistribute the balance between transparency-like reopening and absorptive response in a way that is directly visible in the scattering spectra.
	
	In Model~II, the addition of an internal mediator produces a qualitatively richer structure. After exact elimination of the mediator, the local problem acquires an energy-dependent complex effective coupling between the two qubits. This reorganizes the dressed spectrum into branches with markedly different linewidths and scattering visibility, and makes the distinction between transparency-like and absorption-dominated responses substantially sharper. The resulting passive spectra provide a clearer example of dimerization-sensitive transparency-versus-absorption windows than in the simpler interferometric geometry.
	
	In the active regime, the same Green-function framework makes it possible to analyze how non-Hermitian mode hybridization reorganizes the two-port scattering response. For both models, near-exceptional-point parameter ranges are reflected in the effective Hamiltonian through the coalescence tendency of dressed branches and the suppression of phase rigidity. In Model~II this reorganization is especially transparent: the singular-value valley associated with near-coherent perfect absorption is tied to the narrow quasi-dark side of the dressed spectrum, whereas the largest pole-dominated response is displaced toward the more strongly hybridized bright side. In this sense, the auxiliary mediator does not merely enhance non-Hermitian effects in a uniform way; it redistributes zero-like and pole-like scattering responses among different dressed branches of the same local device.
	
	The broader implication of these results is that an SSH superconducting waveguide should be viewed not just as a passive host for local scatterers, but as a structured environment that can be used to reshape non-Hermitian superconducting microwave functionalities in a controlled way. Because the Green-function reduction separates the role of the extended waveguide from that of the local circuit design, it provides a practical route for comparing distinct device geometries on equal footing and for identifying which functionalities are imposed by topology, which by internal mode hierarchy, and which by active non-Hermitian tuning. We expect this framework to be useful for the design of superconducting microwave devices that exploit structured-waveguide dressing to realize dimerization-sensitive filtering, transparency-versus-absorption switching, and controlled access to near-CPA and pole-dominated operating regimes.
	
	Although the present work focuses on single-photon scattering by small artificial-atom clusters, the Green-function reduction developed here is considerably more general. The same formalism can be extended to multi-photon transport by enlarging the local excitation subspace and can also be adapted to giant-atom waveguide-QED architectures with nonlocal coupling points. In such cases, the SSH waveguide can still be integrated out exactly, while the resulting effective Hamiltonian naturally incorporates the nonlocal interference generated by spatially separated coupling vertices. %This will be the focus of our follow-up investigations. 
	This generalization serves as our immediate research objective and opens avenues for future studies of nonlinear multiphoton dynamics and non-Hermitian topological optics with giant superconducting emitters.
	
	% \textcolor[rgb]{0.7725, 0.3529, 0.0667}{}
	
	\section{Data availability}
	The numerical data and codes that support the findings
	of this work are available from the corresponding author
	upon reasonable request.

	\begin{acknowledgments}
		We acknowledge the support of the National Natural Science Foundation of China (Grants No.	12275193, 11975166) and Science \& Technology Development Fund of Tianjin Education Commission for Higher Education(No. 2024KJ059).
	\end{acknowledgments}
	
	\bibliography{Bibitem}
	%\begin{thebibliography}{99}

\end{document}